\documentclass[12pt]{iopart}

\usepackage{graphicx}

\begin{document}

\title[The colours of Newton's {\it Opticks}: an advanced project for excellent students]{The colours of Newton's {\it Opticks}: \\ an advanced project for excellent students}

\author{Salvatore Esposito}

\address{Istituto Nazionale di Fisica Nucleare, Sezione di Napoli, Complesso Universitario di Monte
S.\,Angelo, via Cinthia, I-80126 Naples, Italy}
\ead{Salvatore.Esposito@na.infn.it}
\vspace{10pt}
%\begin{indented}
%\item[]July 2017
%\end{indented}

\begin{abstract}
We describe in detail an advanced project devised for outstanding High School (or undergraduate) students with appropriate abilities in physical reasoning (rather than with a good standard preparation), centered around the well-known historical case of Newton's theory of light and colours. The different action lines along which the project is developed are aimed to let the students involved to: 1) {\it think} as Newton did, by building step by step all his knowledge and reasoning; 2) {\it work} as Newton did, by performing the whole series of his original experiments with prisms; 3) {\it deduce} as Newton did about the nature of light and colours; 4) {\it present} the results of their activity (including physics demonstrations) to the general public, in order to test abilities in communicating what learned and discovered (including video realization published on YouTube \cite{youtube}). Such didactic aim is complemented by the purpose to realize a historically informed activity, given the potential key role of the History of Physics in promoting science at a deeper level, especially when no particular training in mathematics or advanced education is required. The highly favourable reception of the project by the students involved, as well as that deserved by the uneducated people to the activities demonstrated by the students in public events, testify for the success of such work.
\end{abstract}

% Uncomment for PACS numbers
%\pacs{01.40.E-, 01.50.Fr, 01.50.My, 01.55.+b, 01.65.+g}
%
% Uncomment for keywords
%\vspace{2pc}
%\noindent{\it Keywords}: Science in schools, Newton's theory of colours, experimentum crucis, Science communication
%
% Uncomment for Submitted to journal title message
%\submitto{\JPA}
%
% Uncomment if a separate title page is required
%\maketitle
% 
% For two-column output uncomment the next line and choose [10pt] rather than [12pt] in the \documentclass declaration
%\ioptwocol
%

\section{Introduction}

\noindent The urgency to communicate the basic principles and applications of science, as satisfactorily as possible, to students of any kind and to the general public, has led in recent years to a proliferation of outreach projects and activities of undoubted educational value. A look at the various papers in the present journal, as well as in other specialized ones, is certainly illuminating in this respect. However, the aim of reaching as many students as possible, along with the need of clarity in the given presentations for uneducated people, has often led to sacrifice the historical vision of the given topic -- that is, how people arrived at that given result -- and, also, to develop projects just tailored for common students -- not always very interested (at least, at the beginning) in finding out more and better. As a matter of fact, indeed, projects aimed at excellent students are scarce, and historically informed activities are even rarer.\footnote{With notable exceptions; see, for example, the educational activities and workshops of the {\it Fondazione Scienza e Tecnica} in Florence \cite{FST}, well documented in its own YouTube channel: https://www.youtube.com/user/florencefst.} With the aim of filling this gap, we here present an {\it advanced} project expressly designed for outstanding students of a scientific High School or College, whose skills and curiosity not always are properly addressed, given the known problems with teaching and communicating science to standard classrooms. Apart from the success among the students involved in the activities -- both theoretical and experimental -- performed, suggesting to make this project public in order to allow other people to propose it in other contexts, an unexpected result came after the end of the project, when the same students were asked to prepare historically informed physics demonstrations of what learned for the general public. Indeed, such project was presented at the XXXIX Congress of the Italian Society of the Historians of Physics and Astronomy (SISFA) \cite{SISFA}, and later adopted by this Society in order to present its activities to the general public at the 2019 European Researchers' Night in Naples; further, it was also chosen by the Naples' Unit of the Italian National Institute for Nuclear Physics (I.N.F.N.) for the Science Festival ``Futuro Remoto''. These recognitions then urged us to report about the project proposed, and in the following pages we give a detailed description of it. After a general introduction (including how the students were recruited) to the project proposed, in Sect. 3 we describe its first large part aimed at reconstructing the traditional mindset of a typical scholar of Newton's time, i.e. his {\it philosophical} background pertaining light and colours. The subsequent Section, instead, deals with the second large part of the project, properly addressing the realization of all the relevant experiments performed by Newton about prisms, by starting from an inclusive historical reconstruction of them (which can be viewed on YouTube \cite{youtube}). Finally, in Sect. 5, we present our conclusions and outlooks.

\section{Developing an advanced project}

\noindent A small number of educated students (just 10) were selected among the pupils (aged about 17) of the last years of an Italian scientific High School, with a declared pre-calculus physics curriculum ranging from Mechanics to Optics (and also with a philosophy and history curriculum covering from ancient Greeks to the eighteen century). The selection was performed by asking the students who had registered to it to answer few {\it Fermi questions}; specifically:
\begin{enumerate}
\item Give an estimate of the speed (in km/h) of your hair's growth;
\item How much carbon dioxide do you breathe into the atmosphere each year?
\item Santa Claus\footnote{The selection was performed at the beginning of December, thus justifying this last question.} is preparing to visit all the children of Earth who celebrate Christmas. How fast should he travel on Christmas night?
\end{enumerate}
The aim of the selection was, evidently, not to choose students with a good standard preparation (although almost all of the selected ones did have a good scientific preparation), but, rather, the students who passed the test were those who demonstrated at least an appropriate ability in ``physical" reasoning, just according to the spirit of the {\it Fermi questions} \cite{Weisskopf,Baeyer,Weinstein}.

The basic theme chosen for the project, jokingly named ``The colours of Newton's {\it Opticks}", was the Newtonian theory of colours, as deduced by the English scholar from its most famous experiments on prisms. This is, indeed, a major topic in the history of science, since it revealed for the first time the genius of the young Isaac Newton \cite{Westfall} as a follower of the Galilean experimental method. On the other hand, it also does not require any particular training in mathematics or other advanced education -- which often prevent the development of projects that are more appealing to students -- without diminishing the scope of it. However, since the didactic aim was admittedly not just to present Newton's work in a fascinating way, but rather that of allowing the students involved to {\it think} like Newton, in order to let them (as far as possible) to {\it deduce} the same results as Newton did, the structure of the project was a bit unconventional. It was explicitly (and repeatedly) asked to the students to {\it forget} what they already knew  (or, presumed to know) from other sources, and then to behave like illiterate children (but eager to discover!). In a sense, the project was aimed to train novel Newtons from their infancy to adulthood and, in order to do that, {\it all} their knowledge and reasoning about the central theme had to be built step by step, without taking anything for granted.

Such a very pretentious program was then structured into two major parts, only the second of which dealing with Newton's researches. Indeed, the first part was intended as a means of acquiring the same mindset as any modern science scholar of the western seventeenth century, whose training in ``philosophical" reasoning started from ancient Greek tradition, evolved into Medieval and Renaissance thinking, to finally come to Galilei and his precursors and followers. Of course, for obvious reasons, we limited this long journey into traditional ``philosophical" reasoning just to the stops related to light and colours, although several incursions into mathematics (geometry, as a particular case), philosophy, medicine and even physiology proved instrumental to the purpose. 

The entire project extended for a period of about five-six months, with a scheduled two-hour weekly meeting (typically on Saturday, when no other school lessons were expected). In every meeting of the first part, the students were asked to deal with a given topic, offered to their attention by key natural philosophers of the past who effectively posed and dealt with the given problems. Usually, the teacher present to the meetings (whose role was properly that of a coordinator or a provoker, sometimes a guide) only set the point and encouraged the students to reason along the lines traced by the scholar under examination, who were often helped in their task by repeating simple, basic observations and experiences performed or imagined by that author. Now and then, especially for the Medieval authors, the reading of original texts was crucial in order to let the students to fully appreciate how ``philosophical" reasoning was developing, along with how it was presented to the (educated) readers.

The second part, instead, dealt properly with the work by Newton and, after a short presentation of the problem at hand and its historical contextualization, focussed mainly on the basic experiments performed and described by the English scholar. Here, a reconstruction of the whole series of the experiments was performed (see below), without altering their original sequence, though being often not logically consequential, at least as intended today. The teacher then just read the manuscripts were Newton explained the given experiment (and sometimes explained the sense, given the archaic language employed); afterwards, the students were asked to reproduce that experiment by making recourse to the school lab resources (with the addition of some old prisms of the present author). Their own interpretation of the results of the experiment followed (from time to time stimulated by the teacher), later compared with the original one reported in Newton's papers. This structure applied to the whole set of experiments, finally discussing the conclusions they (and Newton) drew in building Newton's {\it New theory about light and colours}.

\section{On the shoulders of Giants: a reconstruction of Newton's {\it philosophical} background}

\noindent The ambitious goal of reconstructing the traditional mindset of a scholar of the seventeenth century, at least for what pertains to light and colours, in order to let the students to think as Newton did, was certainly the most difficult result to get for the teacher, corresponding to the as well most tedious part of the project for students eager to experiment about physics, and certainly not accustomed to such way of proceeding. Nevertheless, the students readily realized its relevance, according to Newton's own statement that ``if I have seen further it is by standing on the shoulders of Giants".\footnote{Letter from Isaac Newton to Robert Hooke (15 February 1676). Historical Society of Pennsylvania: https://digitallibrary.hsp.org/index.php/Detail/objects/9792.} They were also helped by the informal discussion of the several topics and, especially, by the (often) simple observations and experiences performed without any preparation. In order not to make the description of this part exceedingly long, also to appreciate the didactic development of it without minor distractions, in the following we only report the key points discussed in each meeting.\footnote{Greek theories about light and colours are summarized, for example, in Ref. \cite{Mangio}.}

${}$

\noindent \underline{\it 1. Empedocles of Agrigentum} \\
- The things of this world are characterized not only by form but also by colour; since they are made up of the four fundamental elements, there must be a relationship between the elements that form things and the colours of things themselves. Fire and water, two of the four basic elements, are coloured: white for fire, black for water. \\
- {\footnotesize \underline{Observations.} The sun is fire and produces light that is clear, so white is assigned to fire; the water in the sea presents, instead, as dark, and therefore black is assigned to water.}\footnote{Black and white, as always in ancient Greece, also mean light and dark, respectively.} \\
- All the other colors are generated from the combination of black and white. \\
- {\footnotesize \underline{Observations.} At noon the sun appears white, but at dawn and at sunset it appears of different colours: these are produced by the meeting of the particles of the sun (white) that combine in various proportions with the particles of water (black) existing in the atmosphere in different amounts. Deep in the sea the water appears dark blue and black, but if it is illuminated by the sun (fire) it can appear blue or even white: here, colours result from the combination of water with parts of fire emitted by the sun in different proportions. The rainbow is formed by the light of the sun and the water of the rain: its colours are the result of the combination of particles of fire and water and ultimately of white and black.}\\
- Anatomy of the eye, which is like a lantern on a stormy night, made up of fire (residing in the crystalline lens?) which projects outwards, surrounded by water (the humors that surround the crystalline lens?).\\
- How the white and black parts that generate the colours come to the eye. Objects emit effluences that are emanations that spread through the air and reach the eye; they have certain proportions of particles of the four elements: particles of water, earth, air and fire. The eye, like all materials, has pores, or meanders between the elementary solid parts. We are able to perceive colour precisely because these pores allow the effluences to carry inside the eye information on the colour of the object from which they come. \\
- Reading of Fragment 76c4-d5 from Plato's {\it Meno} \cite{Plato}.
%\footnote{From Plato's {\it Meno} (Fragment 76c4-d5) \cite{Plato}:
%\begin{quote}
%\textsc{Socrates}: So do you talk of `effluences' of things, as Empedocles does? \\
%\textsc{Meno}: Certainly.\\
%\textsc{Socrates}: And `channels' into and through which the effluences pass?\\
%\textsc{Meno}: Absolutely.\\
%\textsc{Socrates}: And some of the effluences fit with some of the channels, while others are
%smaller or larger than them? \\
%\textsc{Meno}: That's so.\\
%\textsc{Socrates}: And do you call something `sight'?\\
%\textsc{Meno}: I do.\\
%\textsc{Socrates}: From this, then, `grasp my meaning', as Pindar put it. Colour is an effluence from surfaces, commensurate with sight, and so perceptible.
%\end{quote}}

${}$

\noindent \underline{\it  2. Aristotle of Stagira} \\
- Every colour originates from the mixture, in different proportions, of white and black, or of light and dark, according to three possible methods. First method (juxtaposition): black and white, in very small quantities, are invisible at a certain distance, so we will see another colour. Second method (superposition): black and white one above the other, as happens when black smoke covers the white sun, which then appears red. Third method (complete mixture), the true cause of colour formation: a completely new substance is produced, in which the original features of the elements survive only in a modified form. \\
- The colour of an object is ``real", while that of a rainbow is ``apparent" (rainbow is not an object). \\
- {\footnotesize \underline{Observations.} Solar and lunar rainbows.} \\
- The rainbow is made up of small drops each of which acts like a small mirror, reflecting only three colours (red, green and blue). \\
- {\footnotesize \underline{Observations.} Simultaneous contrast: red threads on a white background appear different from the same red threads on a black background.} \\
- A mathematical treatment of colors, similar to that of sounds. \\
- Reception of Aristotle's theories in the Middle Ages and in later periods: Robert Grosseteste (1175-1253), Aron Sigfrid Forsius (1560-1624), Francois d'Aguilon (1567-1617), Robert Fludd (1574-1637), Athanasius Kircher (1602-1680).

${}$

\noindent \underline{\it  3. Euclid of Alexandria} \\
- A geometric theory of vision with axioms and theorems. Discussion of the seven postulates. \\
- {\footnotesize \underline{Proofs.} From Ref. \cite{Euclid}.
\\ \indent Proposition I: nothing that is seen is seen at once in its entirety. 
\\ \indent Proposition III: every object seen has a certain limit of distance, and when this is reached it is seen no longer. 
\\ \indent Proposition IV: of equal spaces located upon the same straight line, those seen from a greater distance appear shorter. 
\\ \indent Proposition V: Objects of equal size unequally distant appear unequal and the one lying nearer to the eye always appears larger. 
\\ \indent Proposition XII: objects on lines extending forward, those on the right seem to be inclined toward the left, and those on the left seem to be inclined toward the right. 
\\ \indent Proposition LIV: if, when several objects move at unequal speed, the eye also moves in the same direction, some objects, moving with the same speed as the eye, will seem to stand still, others, moving more slowly, will seem to move in the opposite direction, and others, moving more quickly, will seem to move ahead. 
\\ \indent Proposition LVII: when objects move at equal speed, those more remote seem to move more slowly.}

${}$

\noindent \underline{\it  4. Titus Lucretius Carus} \\
- Light (called {\it lux} or {\it lumen}) is an aggregate of atoms launched into space and filling it all at great speed. \\
- Reading of some passages from Book IV (49-62; 185-190; 246-249) of {\it De Rerum Natura} \cite{Lucretius}. \\
- The colour of bodies depends first on light and then on the order, position and movement of atoms; in fact it changes with the type and direction of the light and therefore it is a secondary effect of the impact of the atoms on the eye.\\
- Reading of some passages from Book II (730-738; 795-798) of {\it De Rerum Natura} \cite{Lucretius}.

${}$

\noindent \underline{\it  5. Claudius Ptolemy} \\
- Results of the experience as a basis for geometric deductions.\\
- Colour is a real property of physical objects and without light it is only potential. The two primary colours are white and black, all the others are various mixtures of these two. \\
- {\footnotesize \underline{Observations.} Rotating a potter's wheel daubed with several colours (Ptolemy's {\it Optics}, Book II, N. 96 in Ref. \cite{Ptolemy}). }

${}$

\noindent \underline{\it  6. Avicenna} \\
- Distinction between {\it lumen} (physical agent of light) and {\it lux} (perceptual effect of light). \\
- The intermediate colours between black and white cannot all be obtained by mixing the two extremes. According to empirical observations (a flame turning red in the presence of smoke), a two-dimensional ordering of colours is proposed, with scales that modulate a given colour along a more or less tonal path (for example, a threefold scale from white to black) \cite{Avicenna}.

${}$

\noindent \underline{\it  7. Alhazen} \\
- {\footnotesize \underline{Observations.} The phenomenon of dazzling.} \\
- Light is a physical entity external to the individual, whose prolonged and direct action can damage eyesight and injure eyes. Then, light comes from the object to the eye of the observer and here it meets visual power. \\
- {\footnotesize \underline{Observations.} Experiments about the rainbow with spheres full of water. The rays of light that pass through the sphere are refracted according to measurable angles; each ray is refracted at a certain angle and produces a certain color with a one-to-one correspondence.} \\
- The rainbow is not produced by reflection (according to Aristotle), but by refraction \cite{Alhazen}.

${}$

\noindent \underline{\it  8. Medieval Scholasticism} \\
- Robert Grosseteste: reading and discussion of the treatise {\it De iride} about the rainbow \cite{Grosse}. \\
- Contributions about the real or apparent nature of colours, including the rainbow, by Roger Bacon (mentioning also Albertus Magnus and Theodoric of Freiberg) \cite{Bacon}.

${}$

\noindent \underline{\it  9. The Renaissance period and Johannes Kepler} \\
- Francesco Maurolico and the geometric theory of the {\it camera obscura} \\
- Giambattista della Porta, the refractive properties of lenses and the anatomy of the eye. \\
- Johannes Kepler about lenses: scientific optics in the treatise {\it Dioptrice} \cite{Giudice}. \\
- Johannes Kepler about the eye as a {\it camera obscura}; the experimental proof of the inverted retinal image by Christoph Scheiner \cite{Kepler}. \\
- {\footnotesize \underline{Observations.} Light rays impinging on spheres filled with liquid converge quite well if they enter from a small opening. Kepler realization that such opening is the pupil in the eye, while the crystalline is suitable for the rays to converge in a single point of the retina, on which the image is formed.} 

${}$

\noindent \underline{\it  10. Mechanicism} \\
- The explanation of Nature in terms of mechanical causality and that of phenomena in terms of matter and movement: the contribution by G. Galilei, P. Gassendi, R. Descartes, W. Charleton and R. Boyle. \\
- Galilei: bodies have few {\it essential} qualities: size, shape, position, movement. All the others are {\it sensitive} or subjective qualities. The primary qualities of bodies interact with sense organs, resulting in sensitive qualities. \\
- Colour is a secondary quality (a sensation) regardless of its source (that is, a physical stimulus), which can be a body, or the light of the rainbow or a prism. \\
- Mechanicist assumption about the origin of colours as due to some form of {\it modification} or alteration of white light (extension of the Scholastic theory of apparent colours to all colours).

${}$ \\

\noindent \underline{\it  11. Ren\'e Descartes} \\
- {\footnotesize \underline{Observations.} Experimenting with a glass ampoule to study the formation of the rainbow. The phenomenon is caused by the refraction that light undergoes in the drops of water suspended in the air.} \\
- {\footnotesize \underline{Observations.} Experiment with a prism: colours are generated by just one refraction (provided that its effect is not destroyed by a second counter refraction on a surface parallel to the first); it is not necessary for the surface to be curved (as in the drops of water), being those of the prism all flat; a shadow or something that limits the light from the prism is required in order the colours be generated by the prism.} \\
- Descartes' mechanical model of solar light modification \cite{Descartes}.

${}$

A further meeting followed, which served as a useful recapitulation of thoughts about light and colours {\it before} Newton \cite{GiudiceNewton}: \\
- Light is a homogeneous entity, not composed, capable of different ``qualities" according to its interaction with matter, but which remains essentially illuminating, with the same essence and the same behavior. Modified by refractions and reflections, light generates the different perceptions of color (``modificationism").

In this meeting, an introduction to Newton's work (and his life) was also presented \cite{Westfall}, especially giving an account of the milestones of his theory about light and colours \cite{GiudiceNewton}:\\
- 1666: colour theory;\\
- 1668: project and construction of a reflecting telescope;\\
- 1669-1671: {\it Lectiones Opticae} \cite{NewtonLectiones}; \\
- 1672: letter to Oldenburg on a {\it New theory about light and colours} \cite{Oldenburg}; \\
- 1704: publication of the treatise on {\it Opticks} \cite{NewtonOpticks}.

\section{The celebrated phenomena of colours: Newton's experiments with prisms}

\noindent To the long ``preparation" reported in the previous section, an equally long series of meetings was then devoted to the realization of all the relevant experiments performed by Newton about prisms, from the simplest ones up to the most famous {\it experimentum crucis}. The choice we have made about the historical reconstruction of the whole series of experiments has been that to include {\it every} experiment concerning the subject and reported by Newton (see below). Thereby, we have included also some apparently spurious experiments, without considering their greater or lesser relevance for the ``final" result. The very aim of the project, indeed, was not to prove some (already known) key result, but rather to experiment just as Newton did, in order to let the students to {\it build} (from nothing, in a sense) that key result without too many facilitations. Again, the presence of the teacher served just to drive the students in a given direction, leaving them free to realize their own ``Newton experiment" with the facilities offered by the hosting physics lab, without further help from the teacher. 

The general strategy adopted was the following:
\begin{enumerate}
\item Reading of a passage from Newton's manuscripts where the given experiment was described, typically including the result of the experiment (but without its interpretation).
\item Performing the experiment by recognizing and using any relevant thing from the lab they need (different prisms, lenses, support rods, etc.), also producing by themselves some required means or other things (coloured cards, perforated panels, darkening devices, etc.).
\item Interpretation of the results achieved in the given experiment.
\item Comparison with the interpretation provided by Newton in his manuscripts and discussion.
\end{enumerate}
As the experiments were completed, a cumulative discussion of the different results achieved was as well added. Note that experiments were not necessarily performed in the laboratory; sometimes they were conducted outside it under the direct sunlight (or even with artificial light sources, though the lab was exposed to the sunlight), or in other places that the students found more appropriate. Indeed, no help came from the teacher to achieve optimal conditions, whose realization -- with the aid of Newton's text -- was completely left to the students (several times no ``optimal" conditions were realized at all). This often led to several repetitions of the same experiment (also employing different conditions) in subsequent meetings, with the obvious consequence that the students very readily realized how difficult was to achieve even an apparently simple experimental result, thus fully appreciating what Newton did. Just to mention a few examples, the students realized that best conditions in almost every experiment were realized when employing direct sunlight with a completely cloudless sky; artificial light produced, instead, remarkable results only when using an additional lens (often modifying the original Newton setup\footnote{The historical reconstruction referred mainly to the reasoning behind the results obtained, rather than to the extremely detailed realization of the given experiments, although Newton's description has been conscientiously followed.}). Also, different prisms (differently sized plastic or glass prisms, also with different aperture angles) have been found to serve better in different experiments. As a rule, regardless of the difficulties encountered, no experiment was undertaken unless the previous one was completed (no cumulative text from Newton's manuscripts was provided to the students).

Newton's original manuscripts we directly employed for the description of the experiments are from Laboratory Notebook (Ms. Add. 3975) and {\it Lectiones Opticae} (Ms. Add. 4002) available at the University of Cambridge Digital Library \cite{UCDL}. For the historical reconstruction leading to the final {\it experimentum crucis}, we mainly follow Ref. \cite{GiudiceNewton}, although a somewhat different approach has been used. We report below the complete list of the experiments performed. The brief description following the short (summarizing) quotation from Newton's original manuscript refers to the best realization of the given experiment, whose final result can be realized from the corresponding figures. The whole realization of all the experiments, instead, can be suitably seen on YouTube \cite{youtube}.

\begin{figure}[t]
\hfill{}{
\begin{tabular}{cc}
\includegraphics[height=3.5cm]{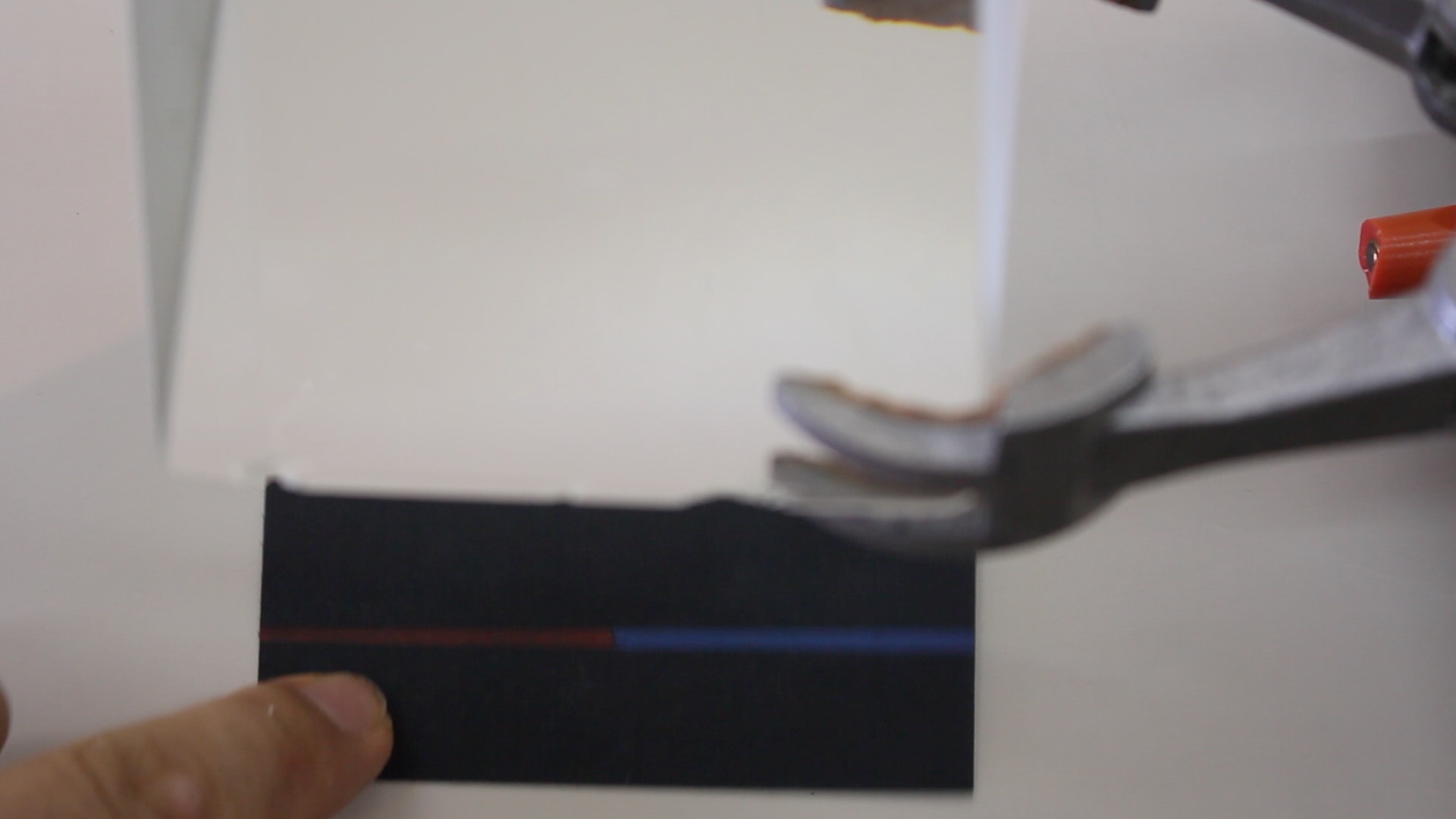} & \includegraphics[height=3.5cm]{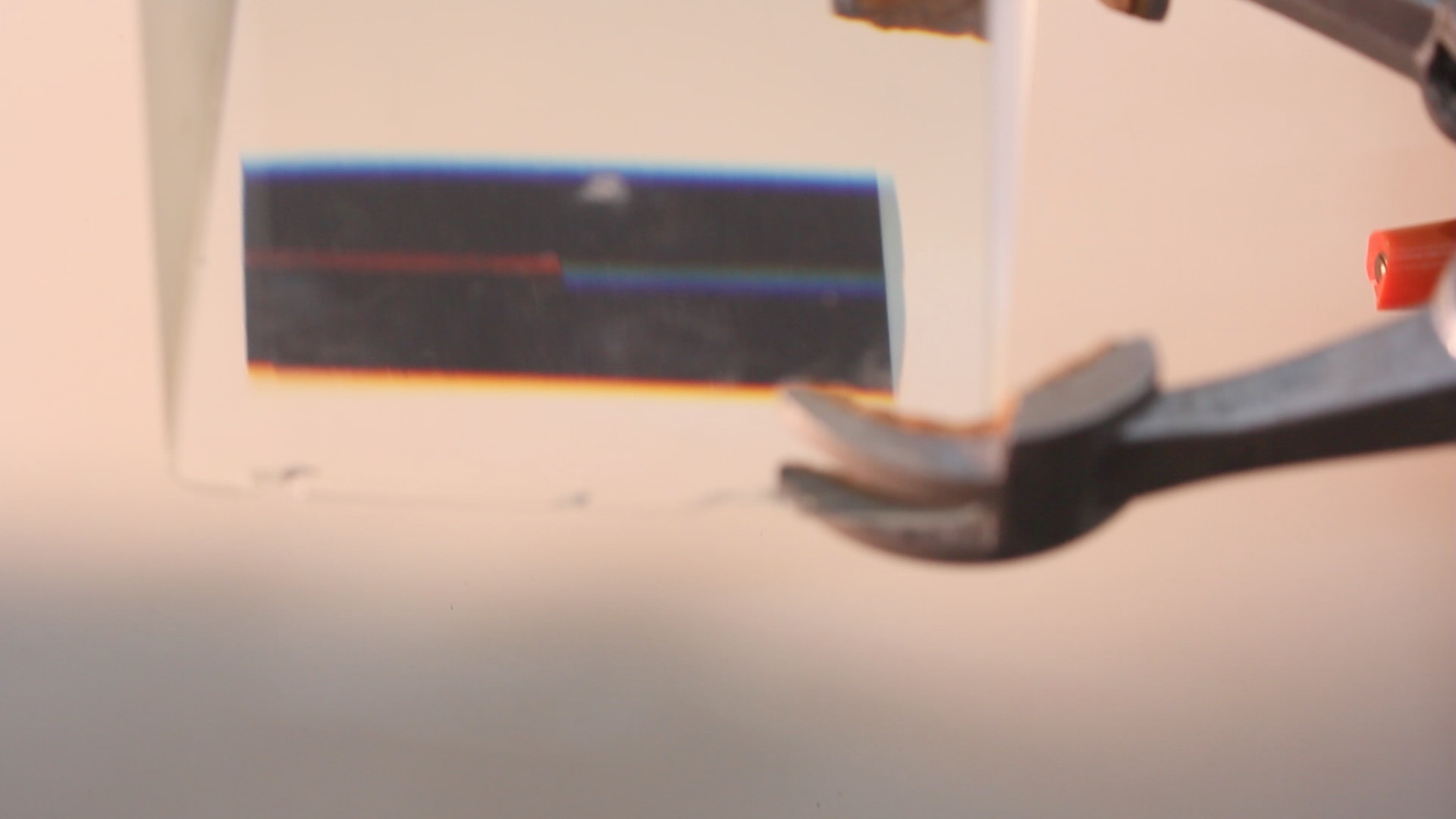} 
\end{tabular} 
\caption{Experiment N.1: a differently coloured line appears broken when observed through a prism.}
\label{fig1}}
\end{figure}

${}$

\noindent \underline{\it Experiment N.1: a broken line}
\begin{quote}
On a black piece of paper I drew a line OPQ, whereof an half OP was a good blew and the other PQ a good red [...]. And looking on it through the prism ADF, it appeared broken in two betwixt colours, the blew part being nearer the vertex of the prism than the red part. So the blew rays suffer a greater refraction than red ones (Ref. \cite{UCDL}, Ms. Add. 3975, page 8).
\end{quote}
The experiment was performed inside the lab, by using a large glass prism (see Fig. \ref{fig1}). The choice of the contrasting colours (red and blue) came out of a careful testing of different couples of colours.

${}$

\begin{figure}[b]
\hfill{}{
\begin{tabular}{cc}
\includegraphics[height=4cm]{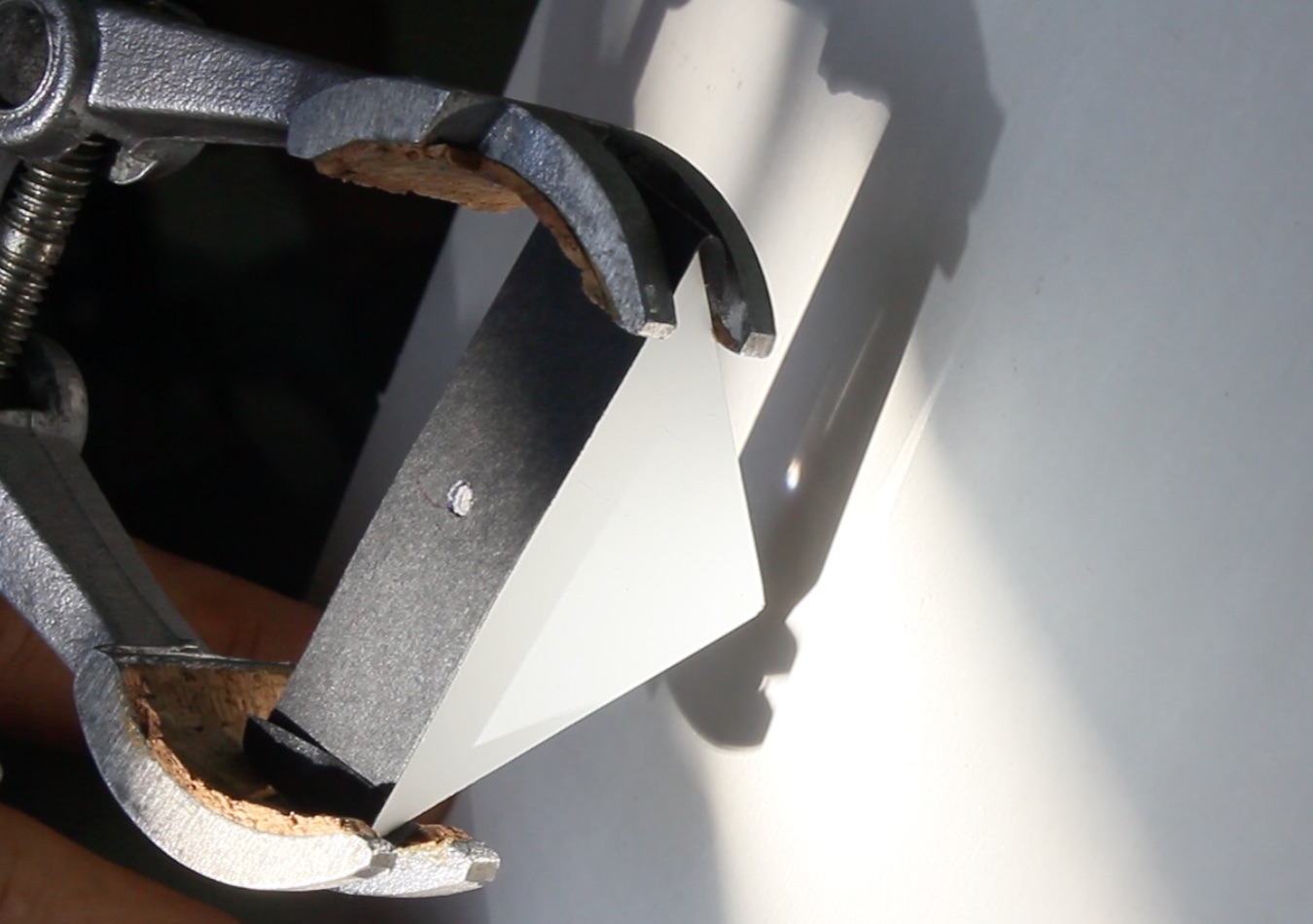} & \includegraphics[height=4cm]{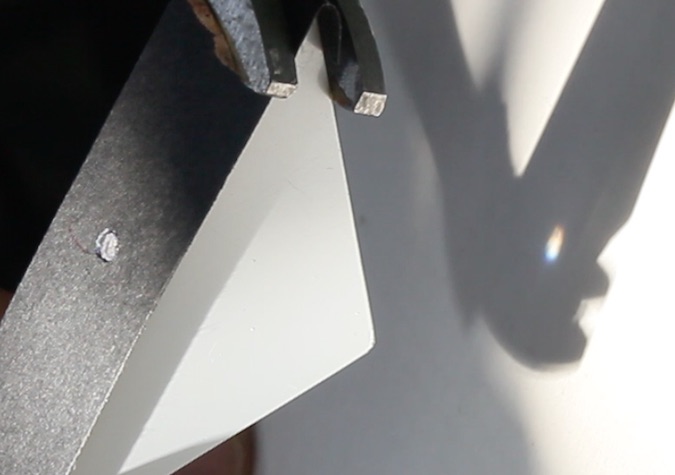} 
\end{tabular} 
\caption{Experiment N.2: an oblong image (with coloured edges) is formed by a prism at a certain distance from it.}
\label{fig2}}
\end{figure}

\noindent \underline{\it Experiment N.2: an unexpected form}
\begin{quote}
Taking a prism into a dark room into which the sun shone only at one little round hole, and laying it close to the hole in such a manner that the rays, being equally refracted at their going in and out of it, cast colours on the opposite wall. The colours should have been in a round circle were all the rays alike refracted, but their form was oblong terminated at their sides with straight lines (Ref. \cite{UCDL}, Ms. Add. 3975, page 8).
\end{quote}
Direct sunlight was directed onto a card with a small circular hole and then passed through a large glass prism. A white screen, upon which the light from the prism was collected, was brought near and far from the prism; the effect was detected at large screen-prism distances, and more efficiently with smaller holes (see Fig. \ref{fig2}).

\begin{figure}[t]
\hfill{}{
\begin{tabular}{cc}
\includegraphics[height=3.5cm]{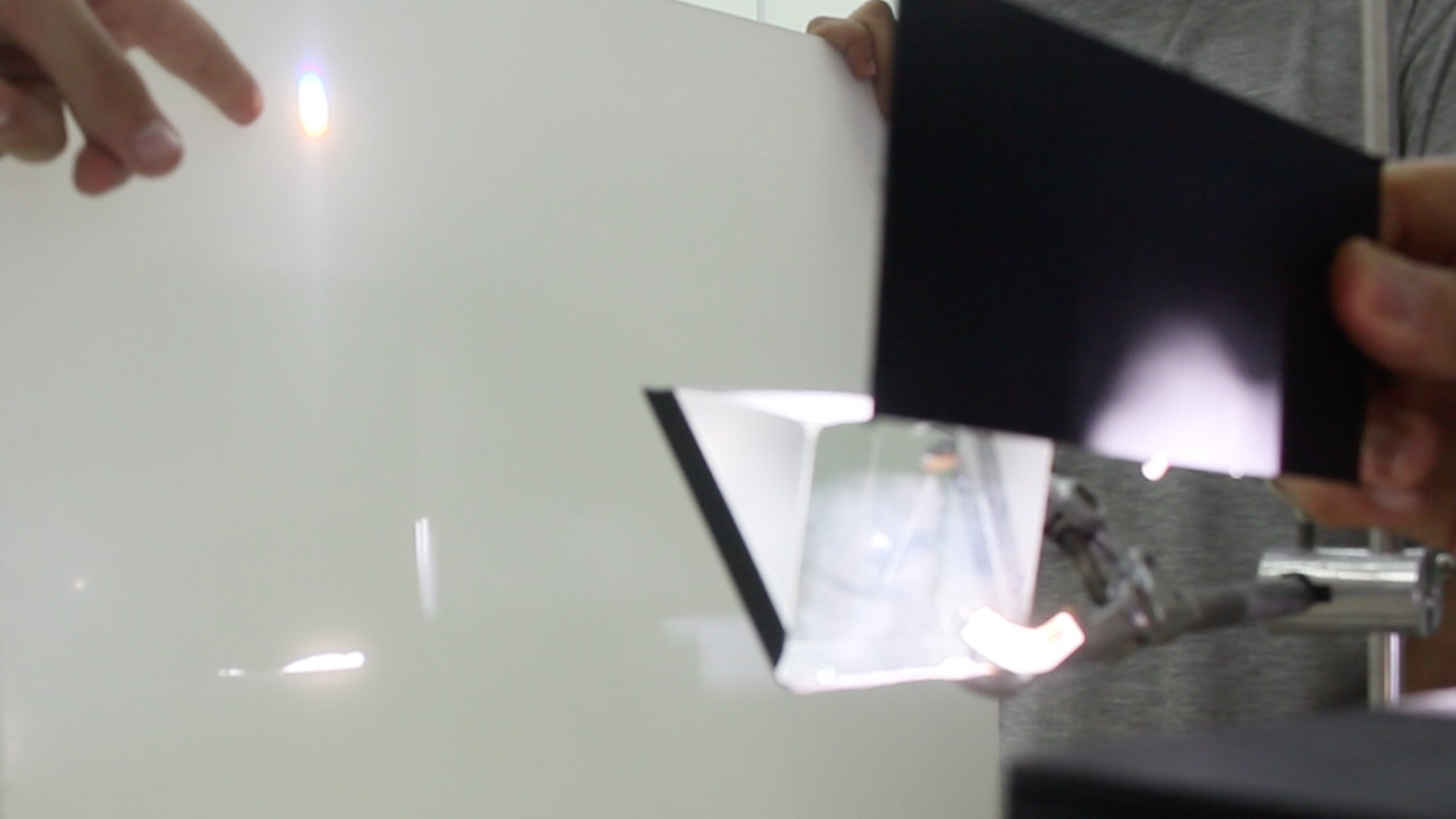} & \includegraphics[height=3.5cm]{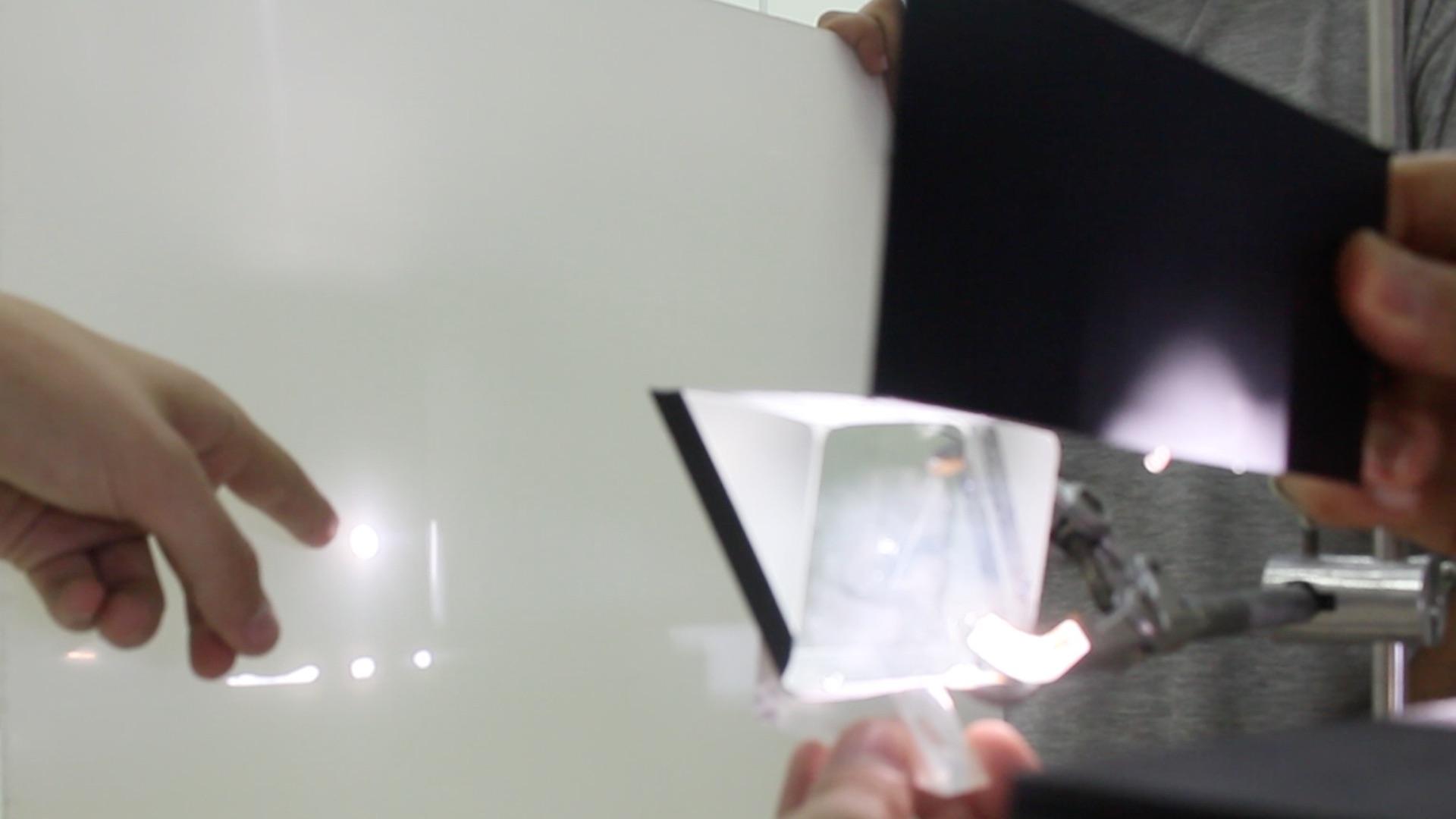} 
\end{tabular} 
\caption{Experiment N.3: a circular image is re-established from the oblong one when a second prism is interposed.}
\label{fig3}}
\end{figure}

${}$

\noindent \underline{\it Experiment N.3: the circle returns}
\begin{quote}
Taking two similar prisms parallel in their longitudes [...] with the sun shining through both in the place Z, where an opaque body is directly set against the light, provided that its rays have first passed through the circular hole F, the light lying in the said place Z will appear distinctly circular, just as if it came directly from F, without the interposition of prisms (Ref. \cite{UCDL}, Ms. Add. 4002, page 29).
\end{quote}
As in the previous experiment, direct sunlight was used. The oblong image came from a large glass prism, while the re-establishing of the circular image was obtained with a smaller plastic prism placed very close to the first prism (see Fig. \ref{fig3}).

${}$

\begin{figure}[b]
\hfill{}{
\begin{tabular}{cccc}
\includegraphics[height=2.8cm]{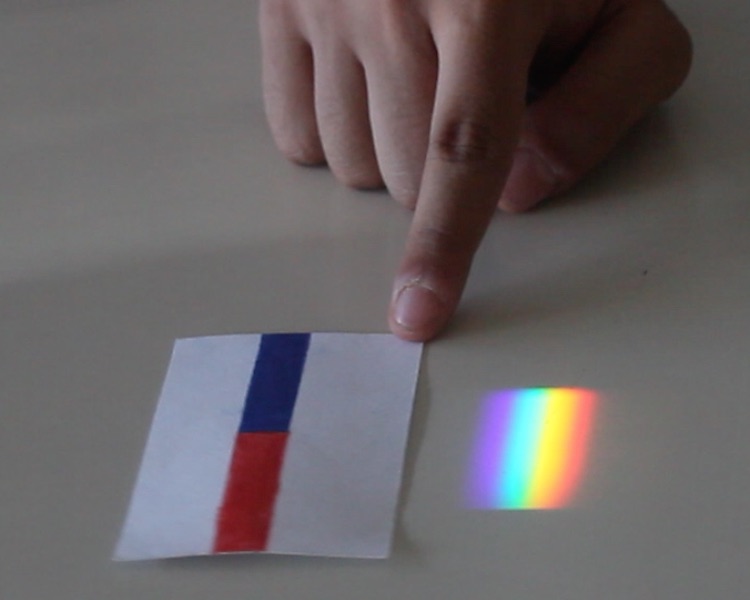} & \includegraphics[height=2.8cm]{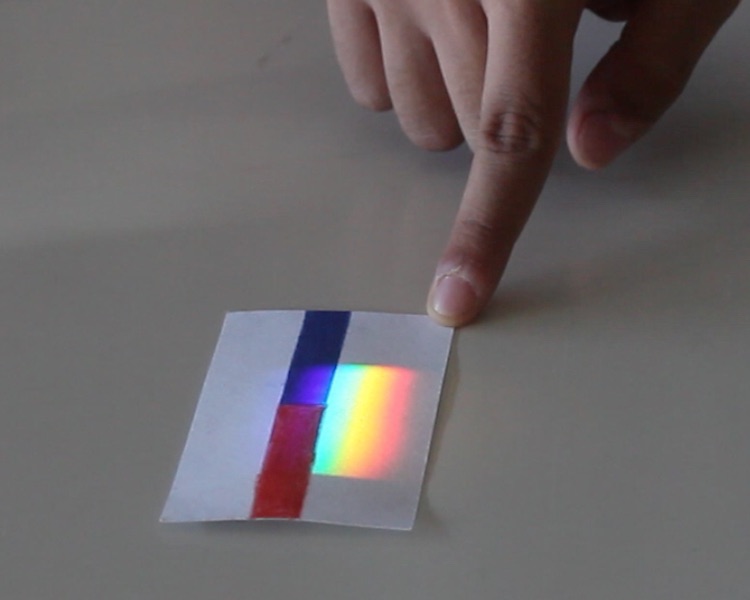} &
\includegraphics[height=2.8cm]{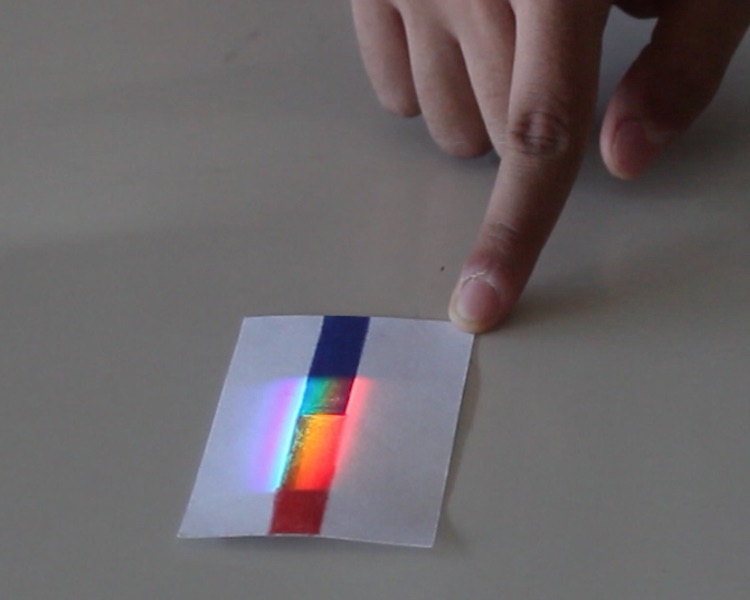} & \includegraphics[height=2.8cm]{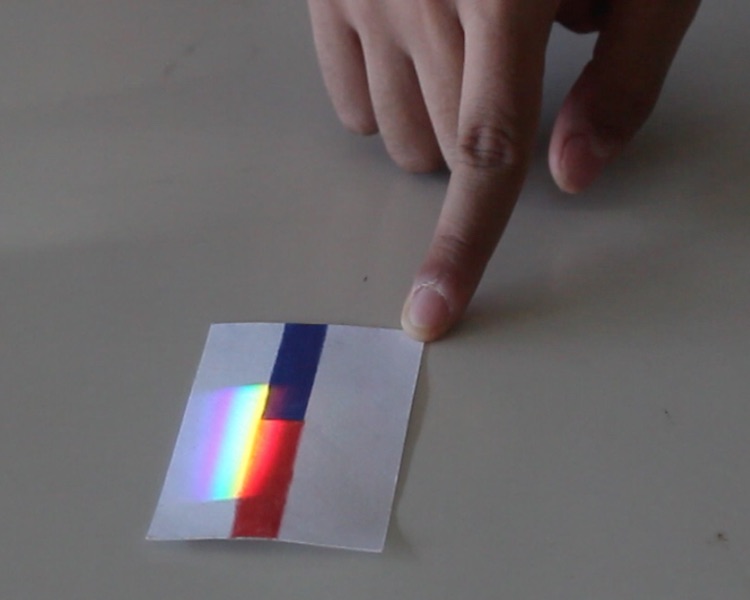} 
\end{tabular} 
\caption{Experiment N.4: red and blue colours appear differently under the rainbow light from a prism.}
\label{fig4}}
\end{figure}

\noindent \underline{\it Experiment N.4: colours over colours}
\begin{quote}
Painting a good blew and red colour on a piece of paper neither of which was much more luminous than the other [...] if the prismatical blew fell upon the colours they both appeared perfectly blew, but the red paint afforded much the fainter and darker blew; but if the prismatical red fell on the colours, they both appeared perfectly red, but the painted blew afforded much the fainter red (Ref. \cite{UCDL}, Ms. Add. 3975, page 9).
\end{quote}
The direct sunlight emerging from a small glass prism was directed through a small slit and then collected on a table. The resulting rainbow light illuminated a red and blue coloured card, very clearly revealing the effect described (see Fig. \ref{fig4}).

${}$

\begin{figure}[t]
\hfill{}{
\begin{tabular}{cccc}
& & \includegraphics[width=3.5cm]{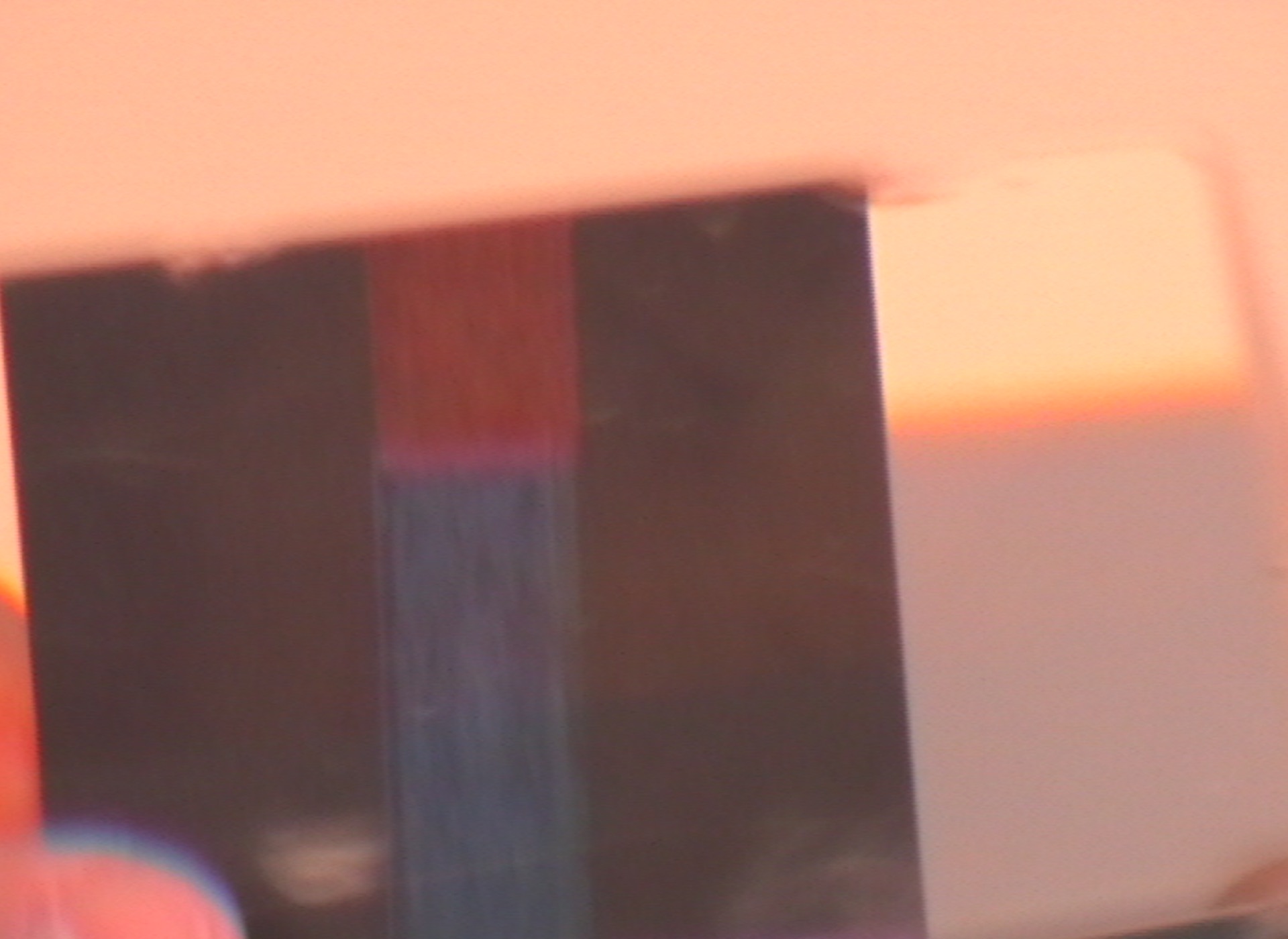} & \includegraphics[width=3.5cm]{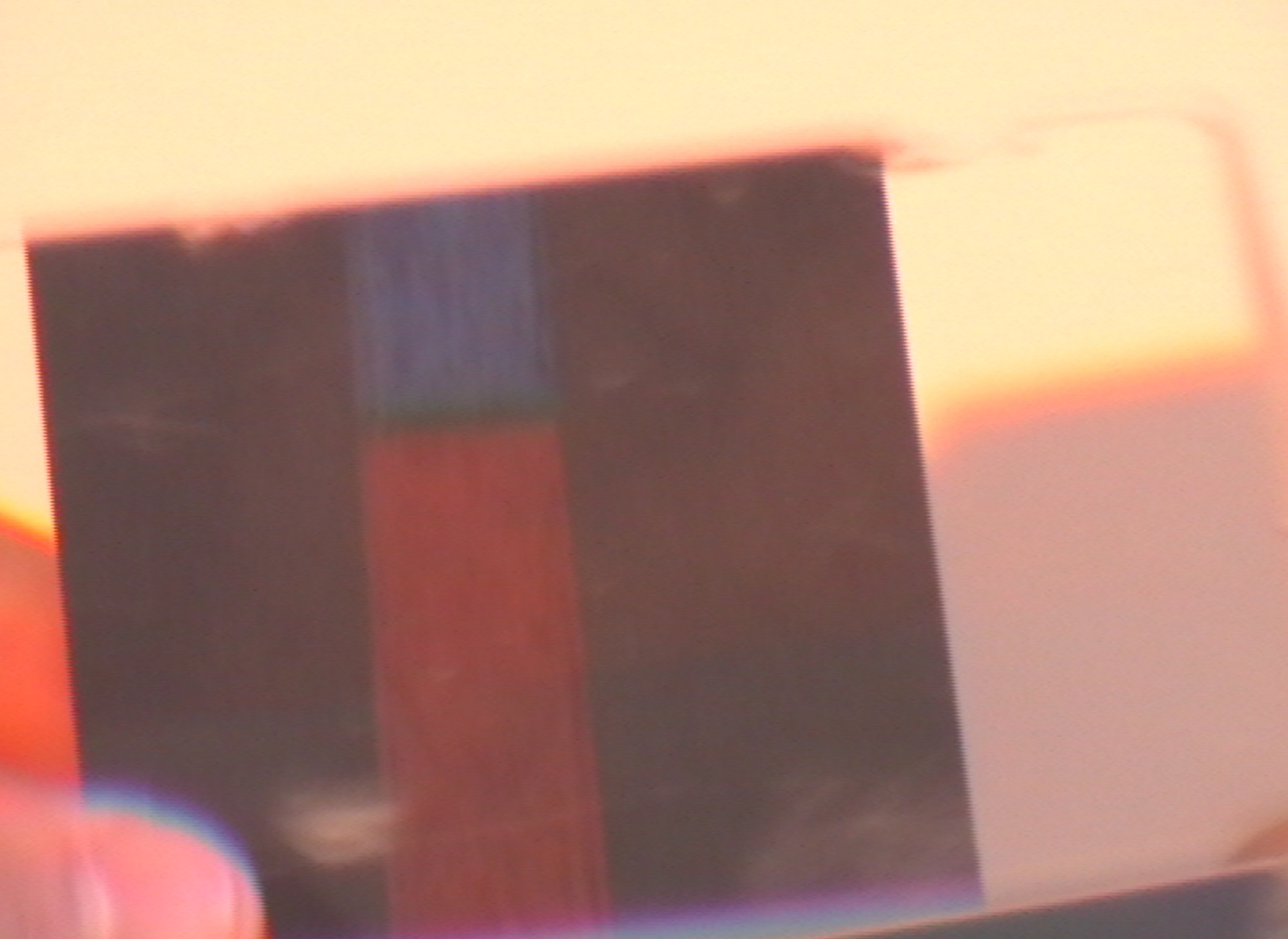} \\
\includegraphics[width=3.5cm]{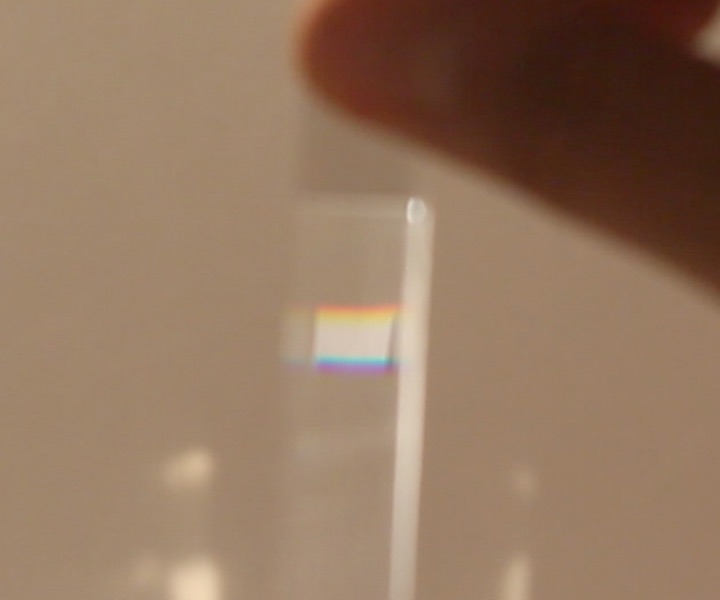} & \includegraphics[width=3.5cm]{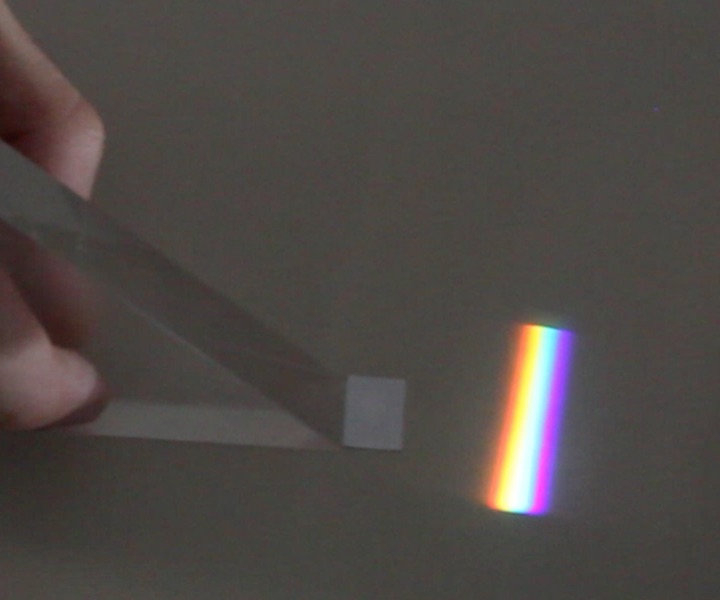} & 
\includegraphics[width=3.5cm]{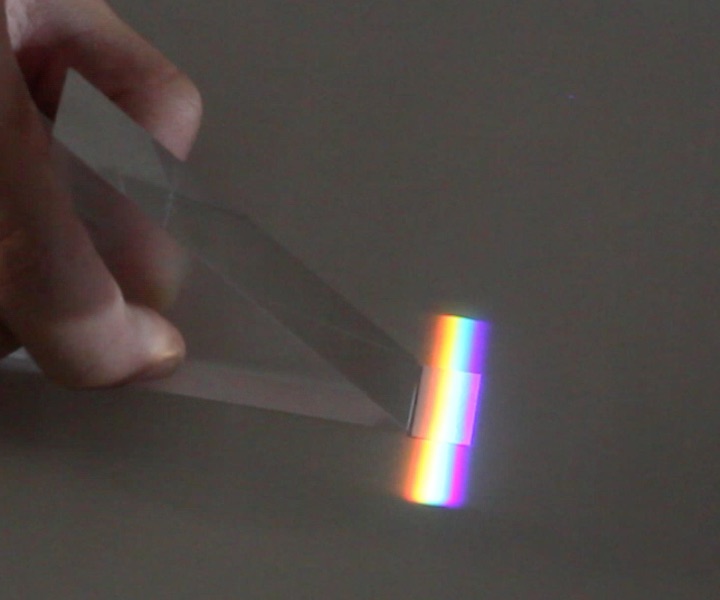} & \includegraphics[width=3.5cm]{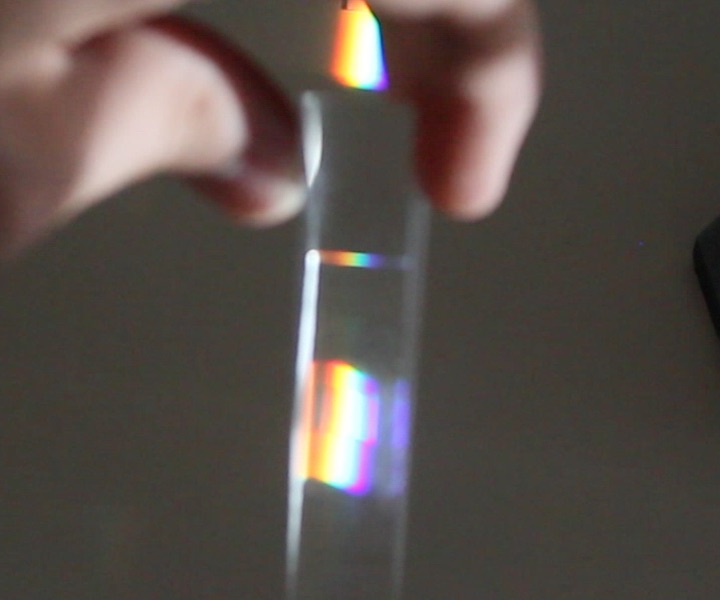} \\
& &  \includegraphics[width=3.5cm]{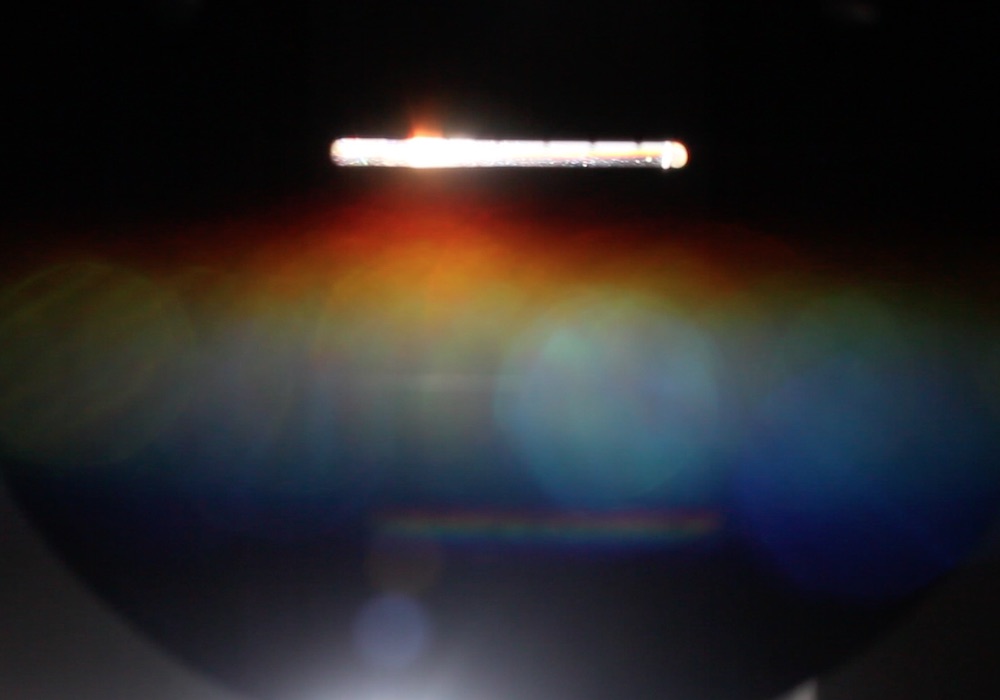} & \includegraphics[width=3.5cm]{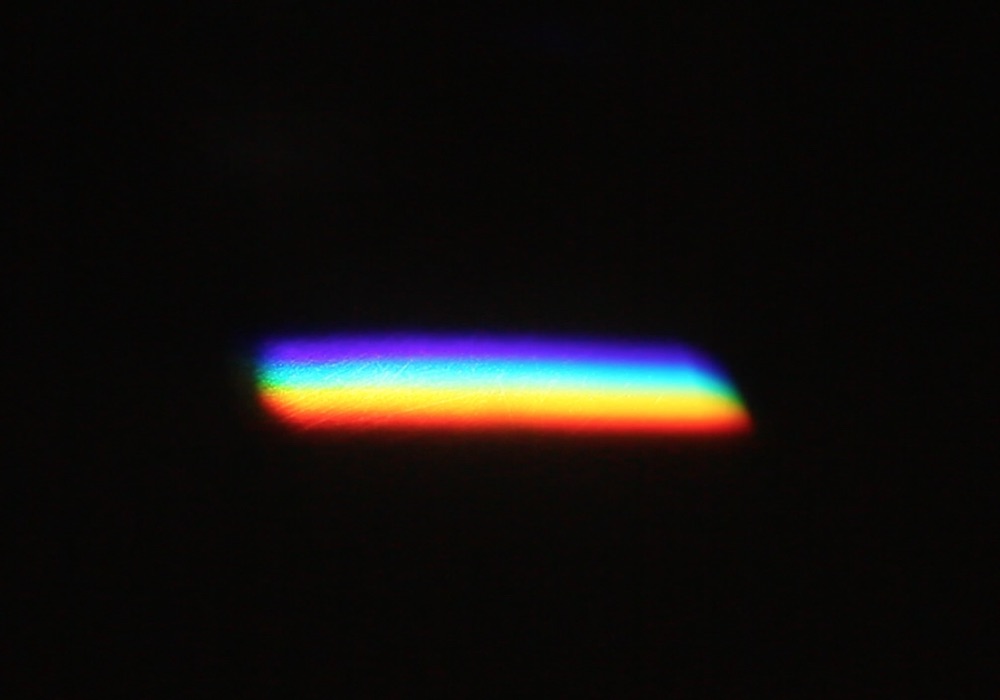} 
\end{tabular} 
\caption{Experiment N.5: the partition edge of a red and blue coloured card appears differently coloured through a prism (first line); the edges of a small sheet of paper appear differently coloured through a prism, while this does not happen when illuminated by the rainbow light from a prism (second line); direct observation (in the eye) of the rainbow light reveals a colour ordering which is different from that observed (reflected) on a screen (third line).}
\label{fig5}}
\end{figure}

\noindent \underline{\it Experiment N.5: partition edges}
\begin{quote}
If the plate ABCDSR be painted with any two colours and ABCD be the lighter colour, the partition edge of the colours will appear through the prism of a red colour, but if CRDS be the lighter colour, their common edge CD will through a prism look blew. [...]\\
But if in a dark room the prismatical blew or red fall on a paper, the edges of the paper will not appear otherwise coloured through another prism than to the naked eye, viz: of the same colour with the rest of the paper.\\
Prismatical colours appear in the eye in a contrary order to that in which they fall on the paper. (Ref. \cite{UCDL}, Ms. Add. 3975, page 10).
\end{quote}
The first part of the experiment was simply realized by looking the partition edge of a red and blue coloured card through a large glass prism. For the second part, instead, a smaller plastic prism was used, while the rainbow light was obtained as in the Experiment N.4. Finally, for the third part of the experiment, direct sunlight passing through a large glass prism and then through a slit was directly observed in the eye and on the table (see Fig. \ref{fig5} where, for the first image in the last line, the rainbow light was directly sent to a camera).

${}$

\begin{figure}[t]
\hfill{}{
\begin{tabular}{cccc}
\includegraphics[width=3.5cm]{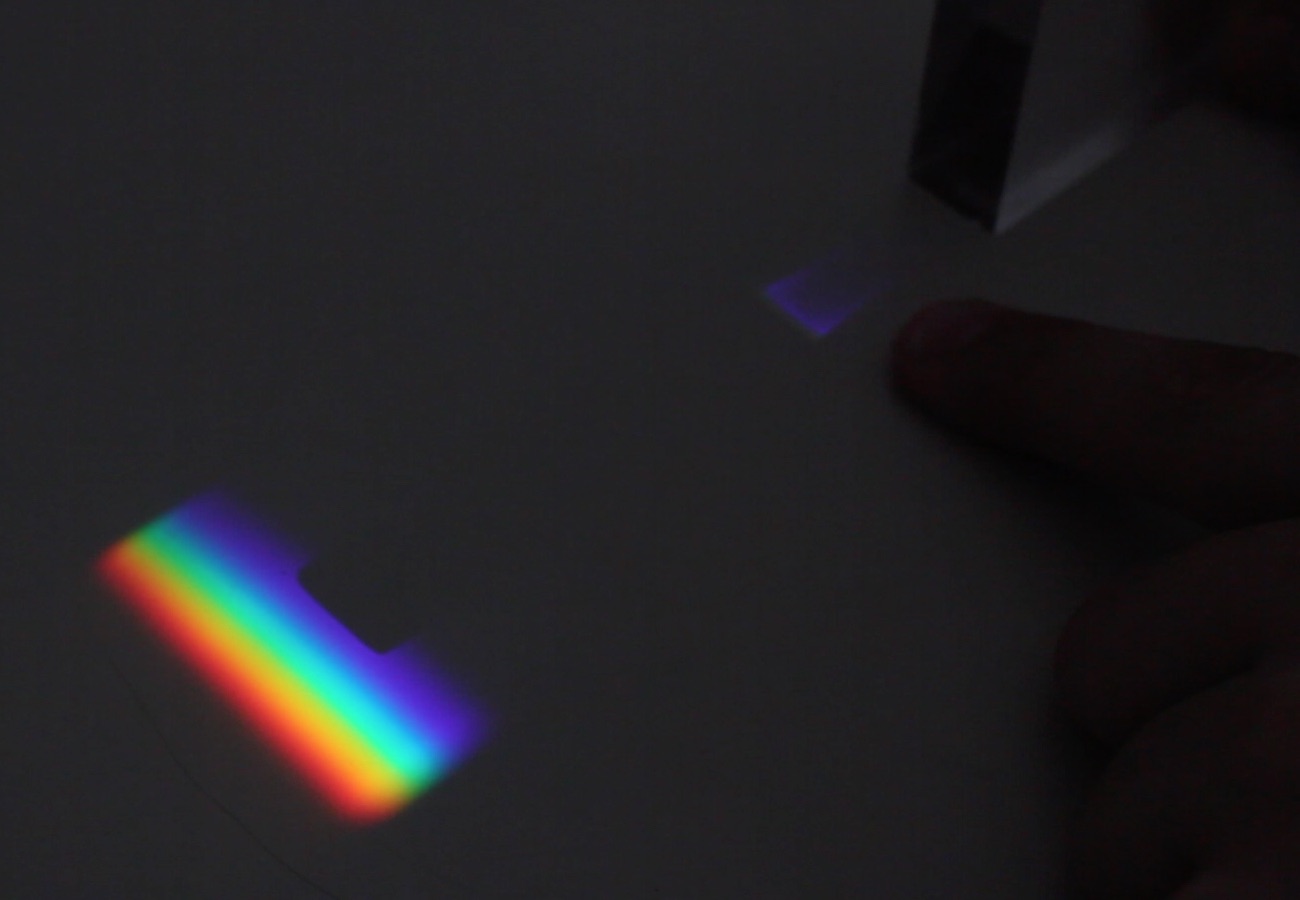} & \includegraphics[width=3.5cm]{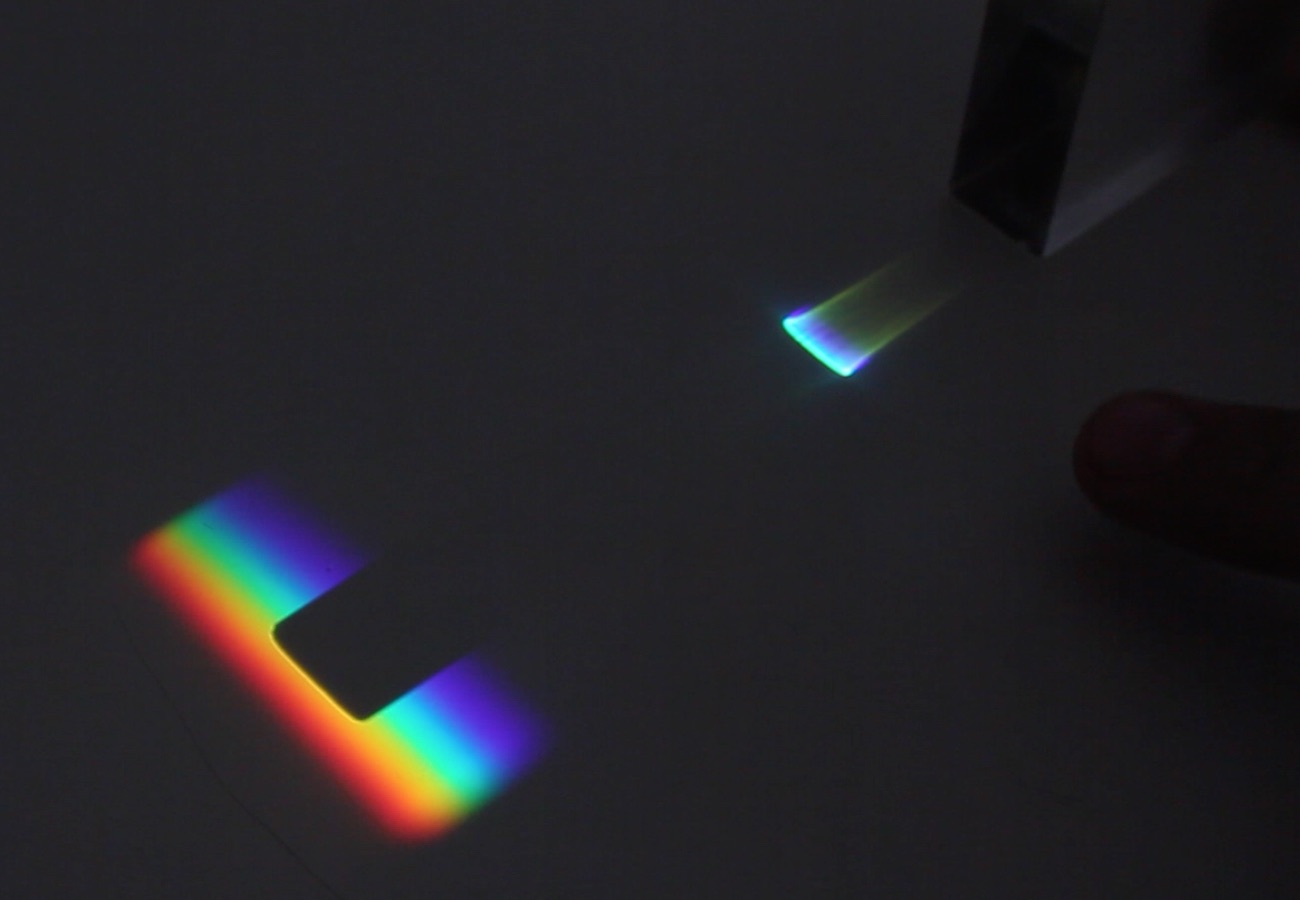} & 
\includegraphics[width=3.5cm]{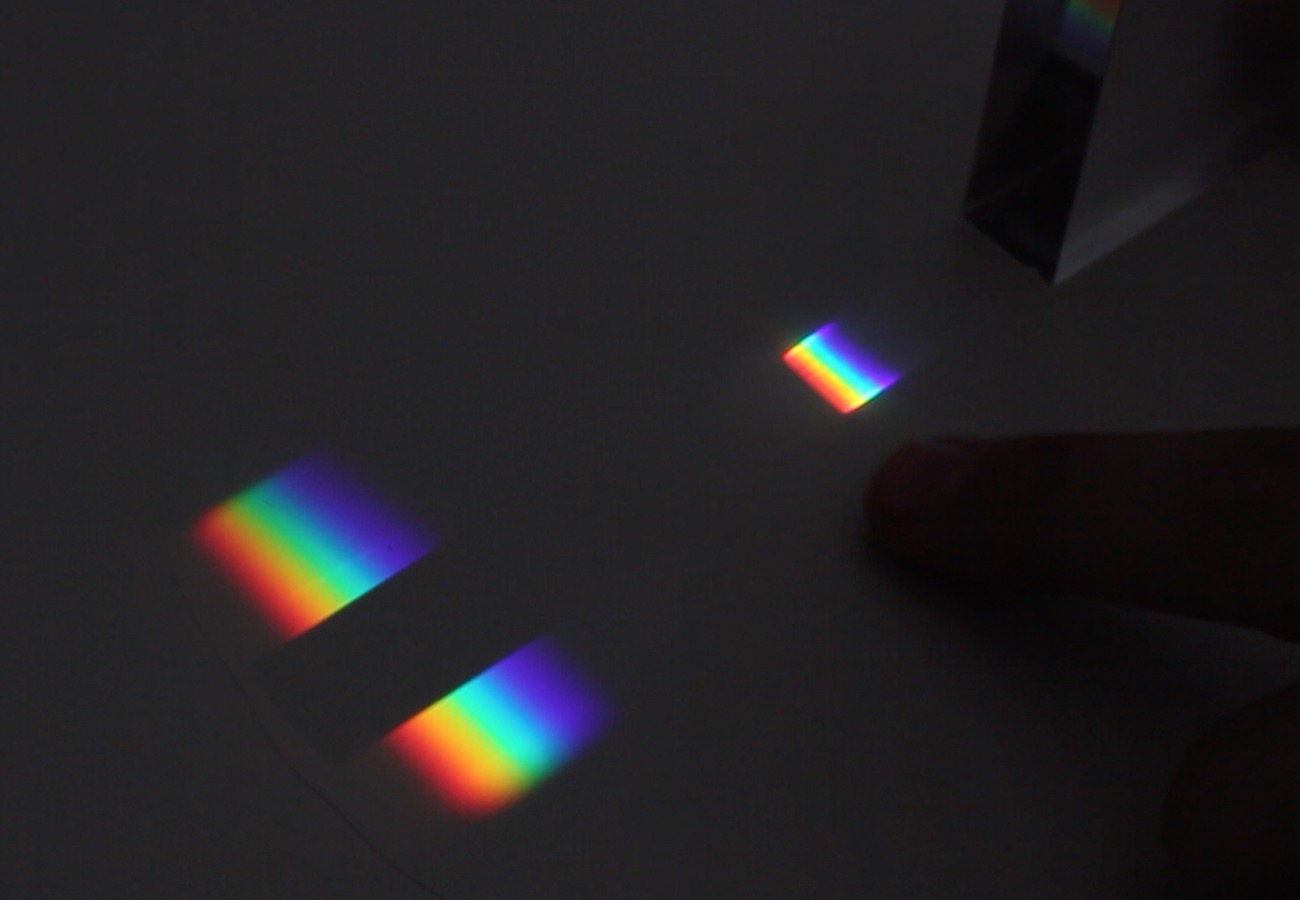} & \includegraphics[width=3.5cm]{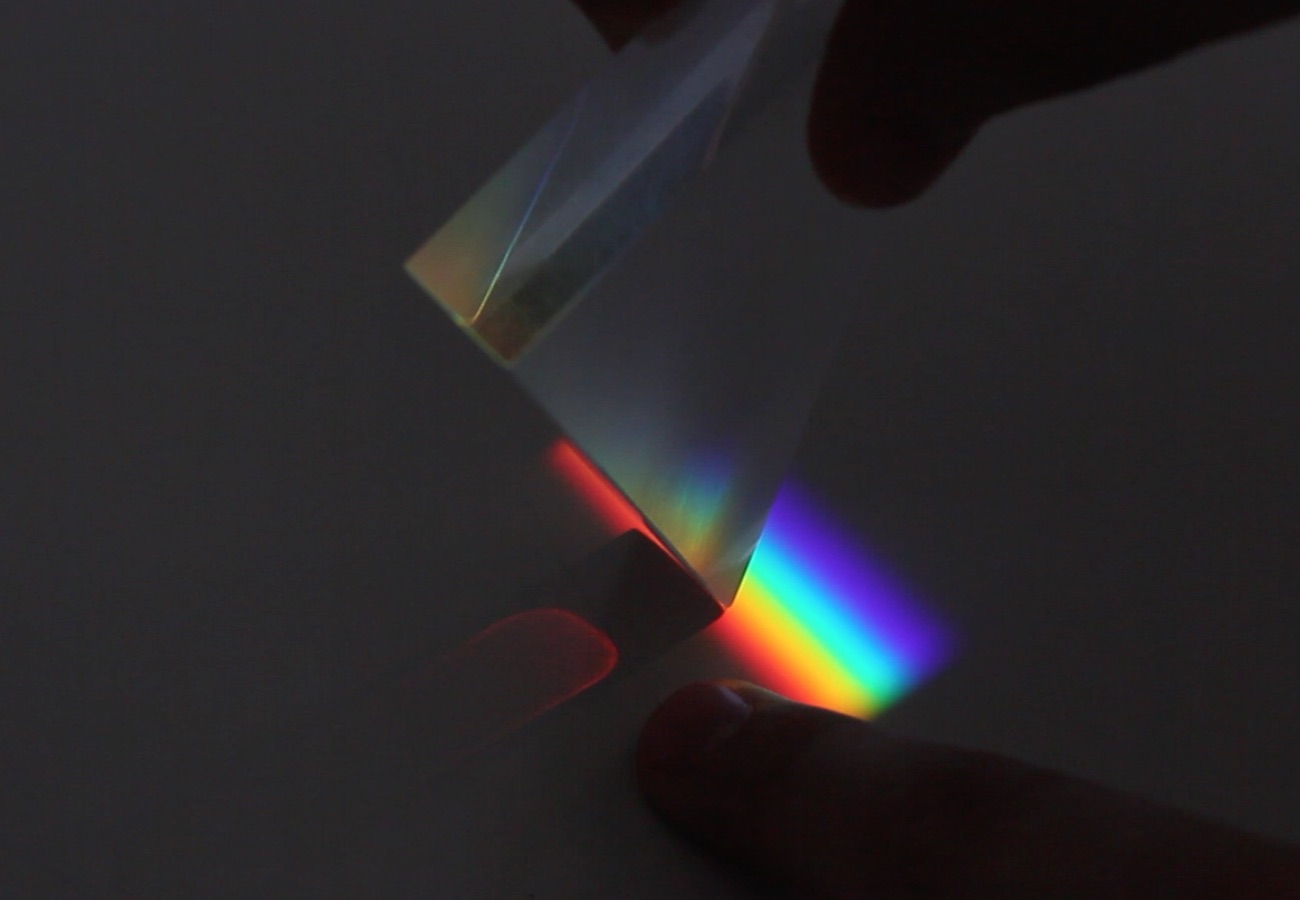} 
\end{tabular} 
\caption{Experiment N.6: coloured lights from a first prism produce no more colours in a second prism, which only shifts their position.}
\label{fig6}}
\end{figure}

\noindent \underline{\it Experiment N.6: a double refraction}
\begin{quote}
Refracting the rays through a prism into a dark room and holding another prism about 5 or 6 yards from the former to refract the rays again, I found first that the blew rays did suffer a greater refraction by the second prism than the red ones. And secondly that the purely red rays refracted by the second prism made no other colours but red and the purely blew ones no other colours but blew ones (Ref. \cite{UCDL}, Ms. Add. 3975, page 18).
\end{quote}
The colours produced by sunlight impinging on a large glass prism and then passing through a slit are allowed to pass (one at a time, i.e. first purple, then blue and so on up to red) through a second, plastic prism. Cumulatively, the colours passing through the second, smaller prism are shifted with respect to the other ones, as clearly visible in the second and third images of Fig. \ref{fig6}. The effect is very striking in a dark room.

${}$

\begin{figure}[b]
\hfill{}{
\begin{tabular}{cccc}
\includegraphics[width=3.5cm]{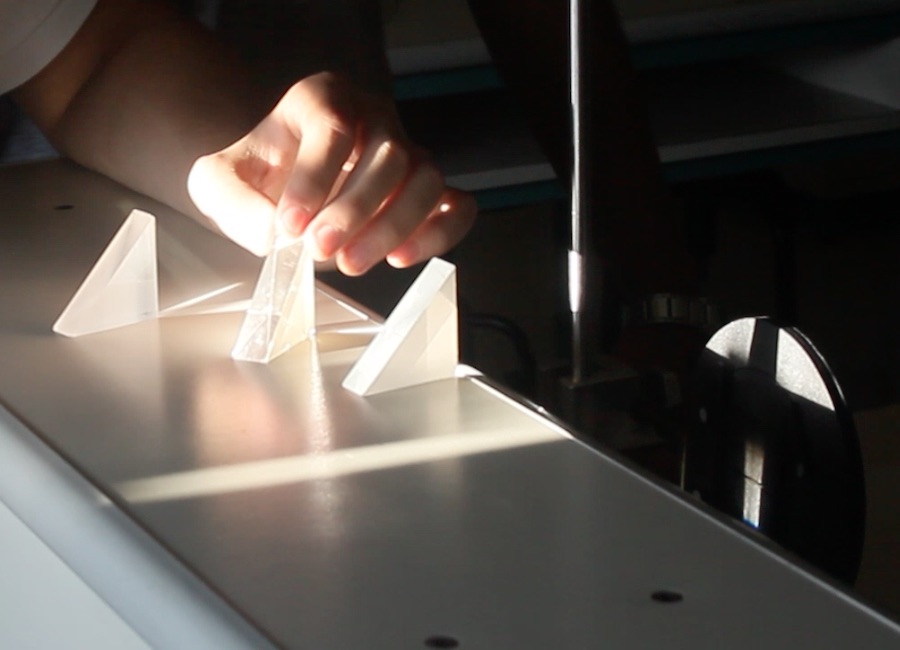} & \includegraphics[width=3.5cm]{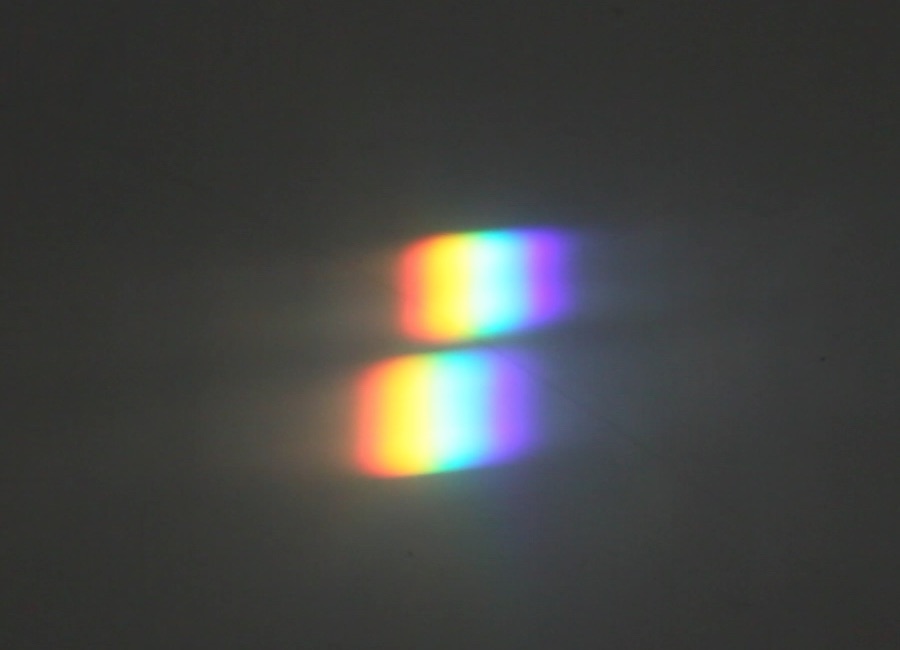} & 
\includegraphics[width=3.5cm]{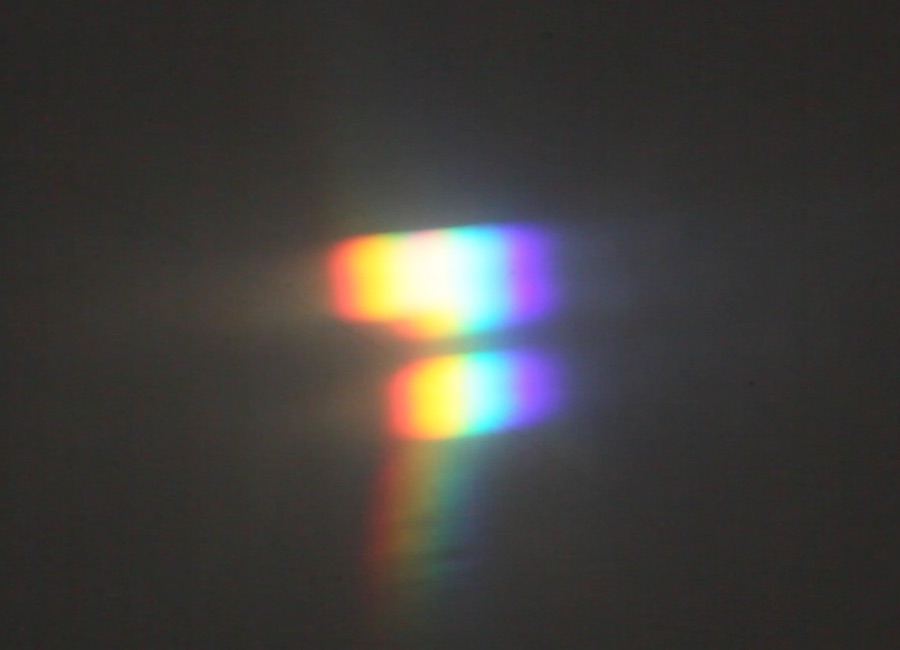} & \includegraphics[width=3.5cm]{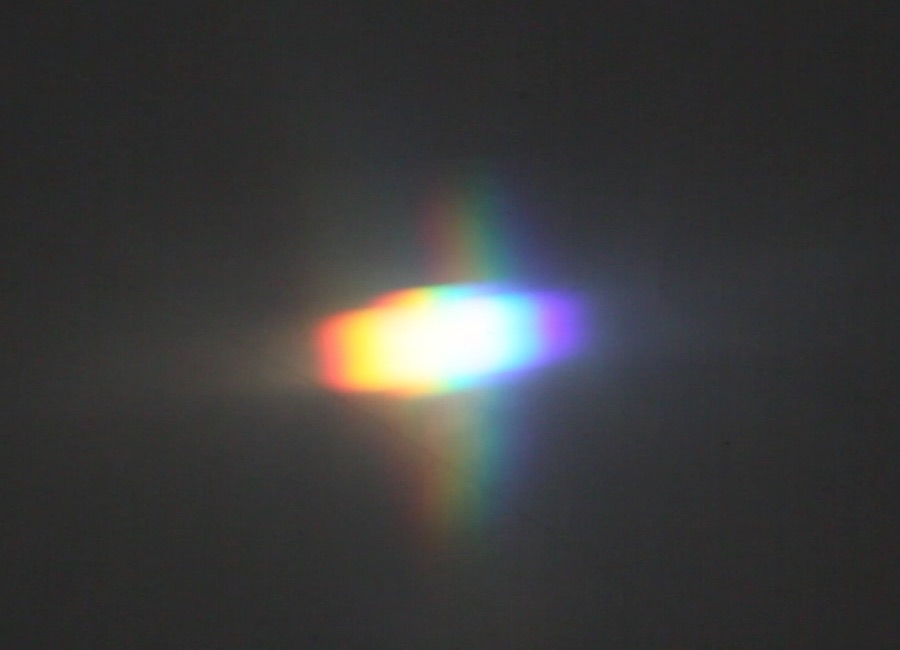} 
\end{tabular} 
\caption{Experiment N.7: different colours from different prisms produce white when they overlap.}
\label{fig7}}
\end{figure}

\noindent \underline{\it Experiment N.7: overlapping refractions}
\begin{quote}
If three or more prisms A,B,C be held in the sun so that the red colour of the prism B falls upon the green or yellow colour of the prism A and the red colour of the prism C falls on the green or yellow colour of the prism B; the said colours falling upon the paper DE at P,Q,R,S. There will appear a red colour at P and a blew one at S but betwixt Q and R where reds, yellows, greens, blews, and purples of the several prisms are blended together, there appears a white (Ref. \cite{UCDL}, Ms. Add. 3975, page 18).
\end{quote}
The direct sunlight passing through three plastic prisms and then through a slit produced three different sets of colours, which were allowed to overlap on a table. The overlapping region (with two or even three sets of colours) showed to be white, while colours retained their nature when they didn't overlap (see Fig. \ref{fig7}). Difficulties in realizing the experiment (with clearer results in a dark room) have essentially been encountered only in properly positioning the prisms to let the colours to overlap; smaller prisms (in somewhat limited space) revealed to be better for this aim than larger ones.

${}$

\begin{figure}[t]
\hfill{}{
\begin{tabular}{ccccc}
\includegraphics[width=2.7cm]{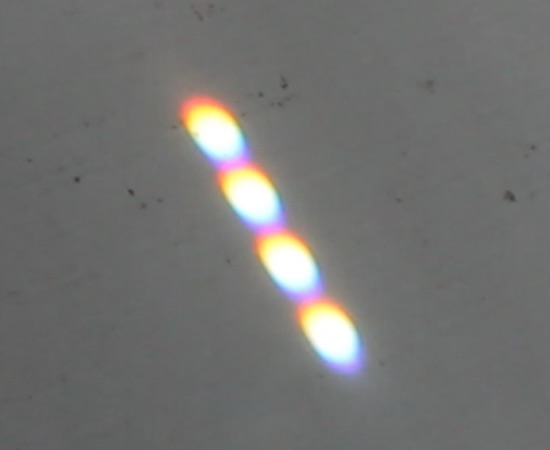} & \includegraphics[width=2.7cm]{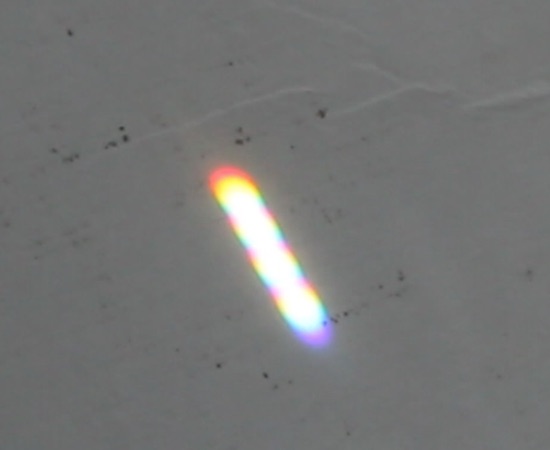} & 
\includegraphics[width=2.7cm]{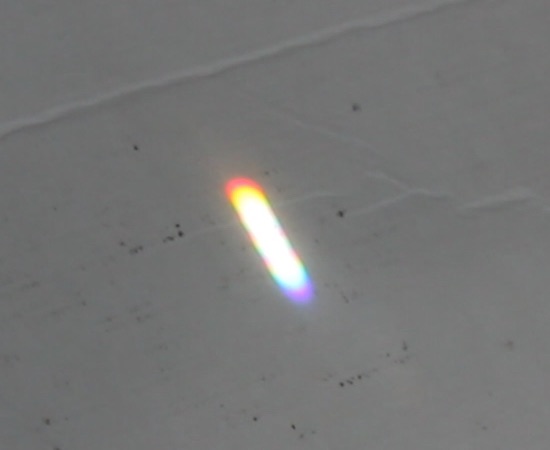} & \includegraphics[width=2.7cm]{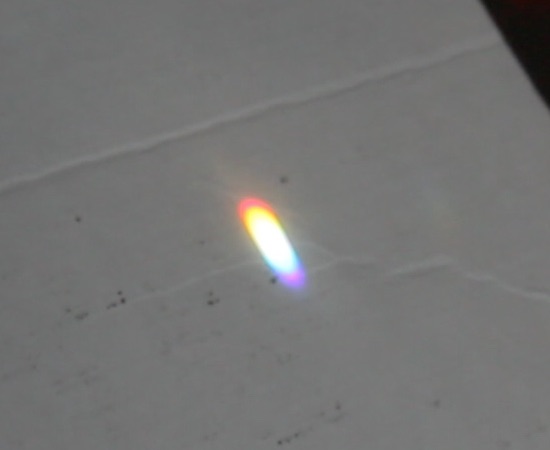} & 
\includegraphics[width=2.7cm]{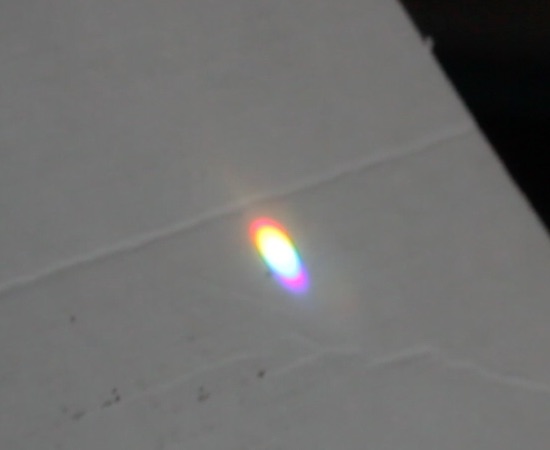}
\end{tabular} 
\caption{Experiment N.8: coloured, oblong images from several holes fuse together in just one white image with coloured edges when the screen moves away from the prism.}
\label{fig8}}
\end{figure}

\noindent \underline{\it Experiment N.8: one prism, many refractions}
\begin{quote}
If you clean a piece of paper on one side of the prism with severall slits [...] so that the light passing through those slits make colours on the paper DE; if the said paper be held near to the prism there will appear for each slit a coloured line. The paper being held farther of until the said coloured lines be blended together, there will appear white where those colours are blended (Ref. \cite{UCDL}, Ms. Add. 3975, page 19).
\end{quote}
Direct sunlight was directed onto a card with several small circular holes and then passed through a large glass prism. A white screen collecting the light was brought near and far from the prism, the effect occurring at large screen-prism distances, and more efficiently with smaller holes (see Fig. \ref{fig8}).

${}$

\begin{figure}[t]
\hfill{}{
\begin{tabular}{cc}
\includegraphics[height=3.5cm]{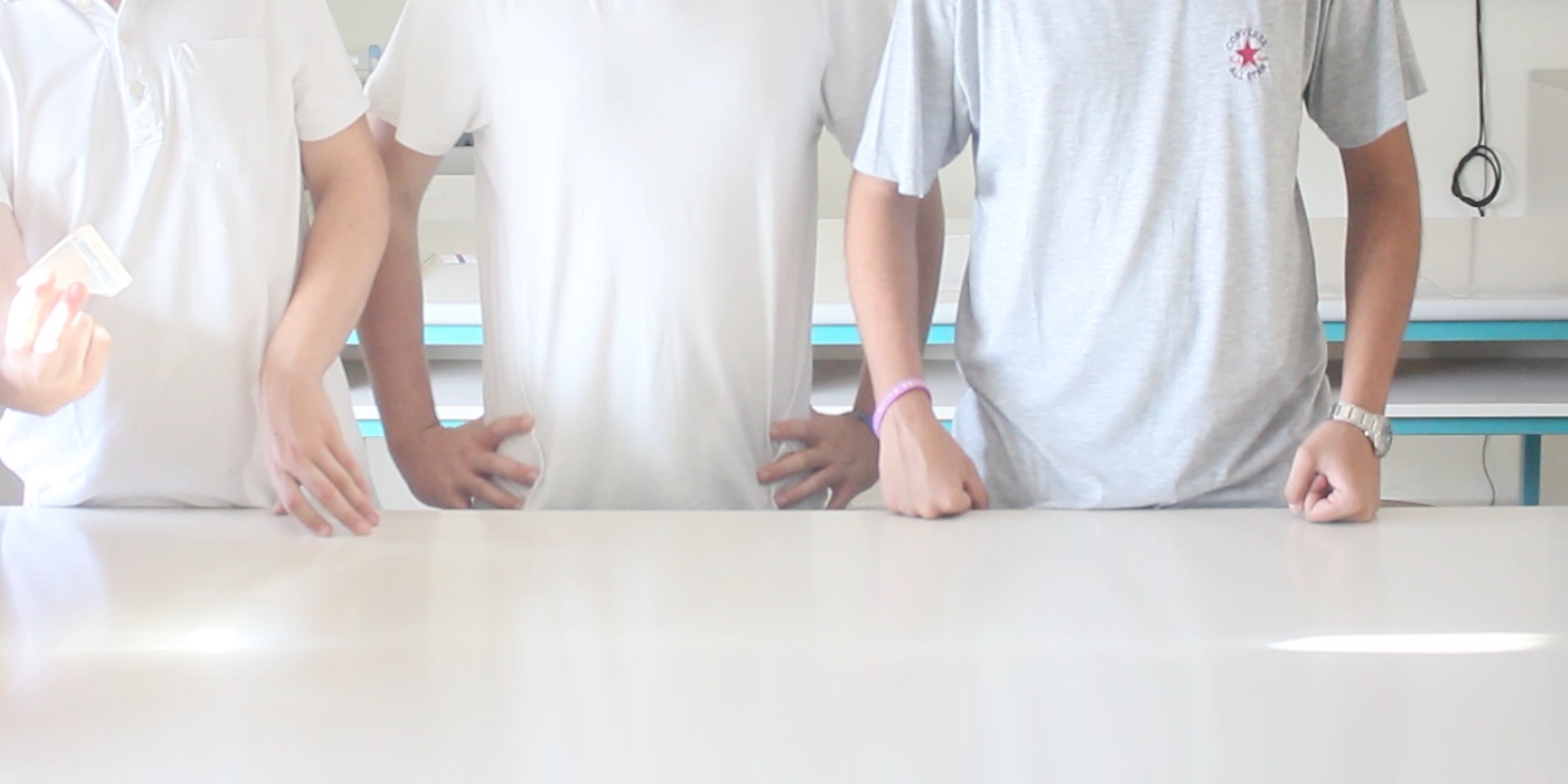} & \includegraphics[height=3.5cm]{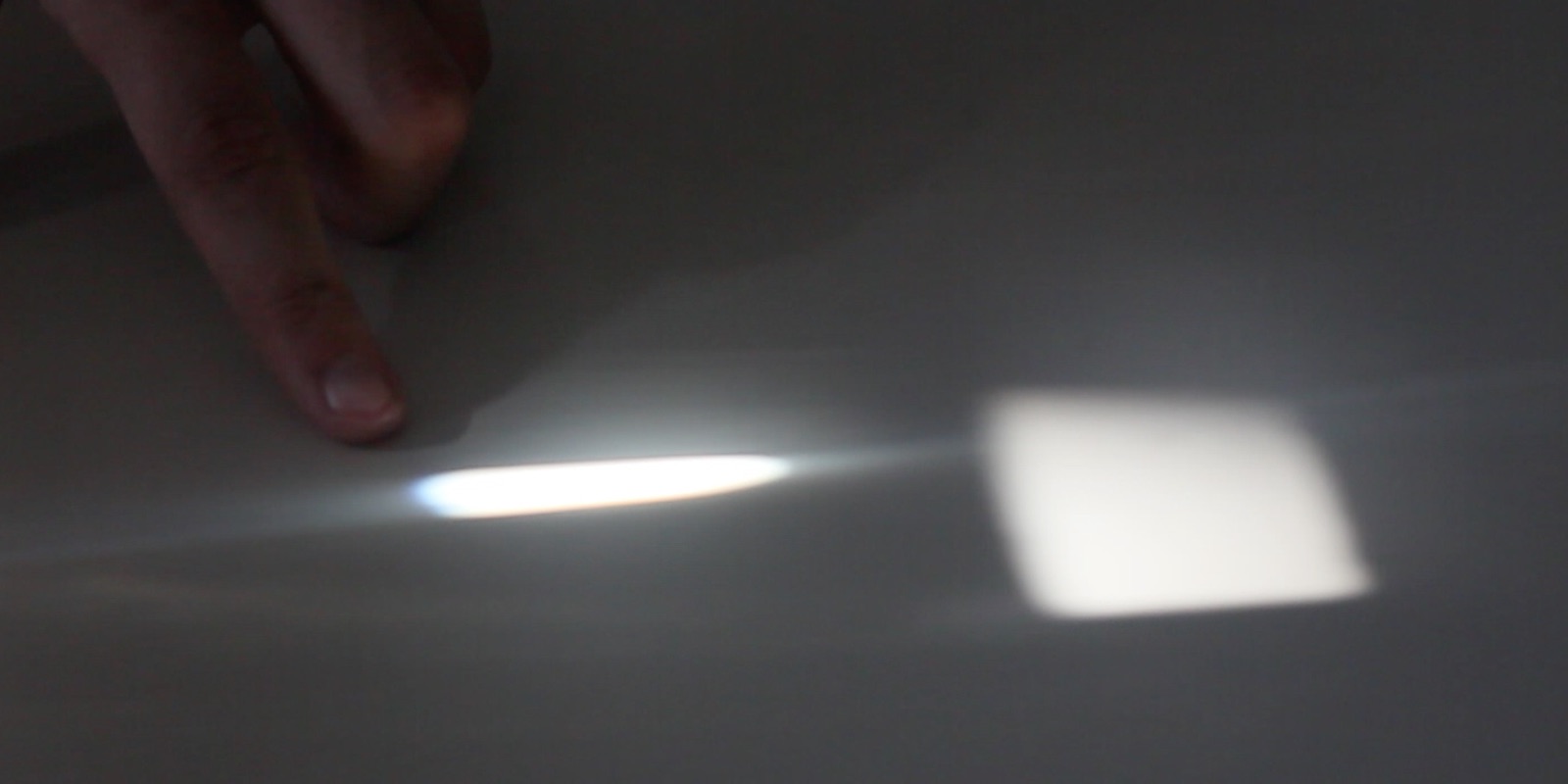} 
\end{tabular} 
\caption{Experiment N.9: when a prism is rotated around its axis, coloured, refracted light gives way to white, reflected light; further refraction following starts with a blue halo replacing white light.}
\label{fig9}}
\end{figure}

\noindent \underline{\it Experiment N.9: reflected vs refracted}
\begin{quote}
The surface of the BC prism does not refract all the rays towards T but also reflects many in P [...]. Rotate the prism on its axis according to the order of the letters ABCA, and you will see both the extension of the colours towards T and the amount of light towards P continually increase [...], until the colours in T begin to fade and are reflected towards P. [...] You will see the white in P change little by little and tend somewhat towards the blew, due to the entrance of the purple and blew that are reflected first; but after the other colours are also reflected by T, white will be restored in P (Ref. \cite{UCDL}, Ms. Add. 4002, page 60).
\end{quote}
This quite difficult experiment was better realized just by rotating the large glass prism (illuminated by sunlight) with one hand, and collecting the reflected/refracted light on a table. The blue halo is quite clearly visible in a darkened room (see Fig. \ref{fig9}).

${}$

\begin{figure}[t]
\hfill{}{
\begin{tabular}{ccc}
\includegraphics[width=4.7cm]{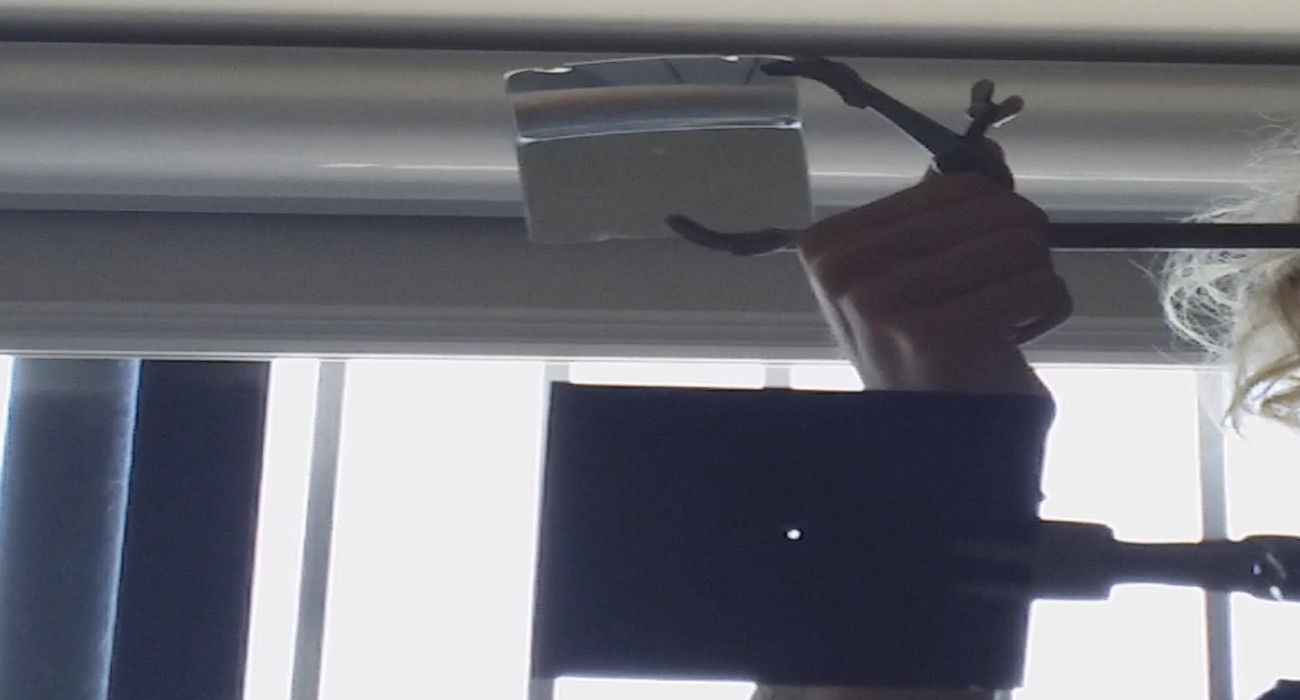} & \includegraphics[width=4.7cm]{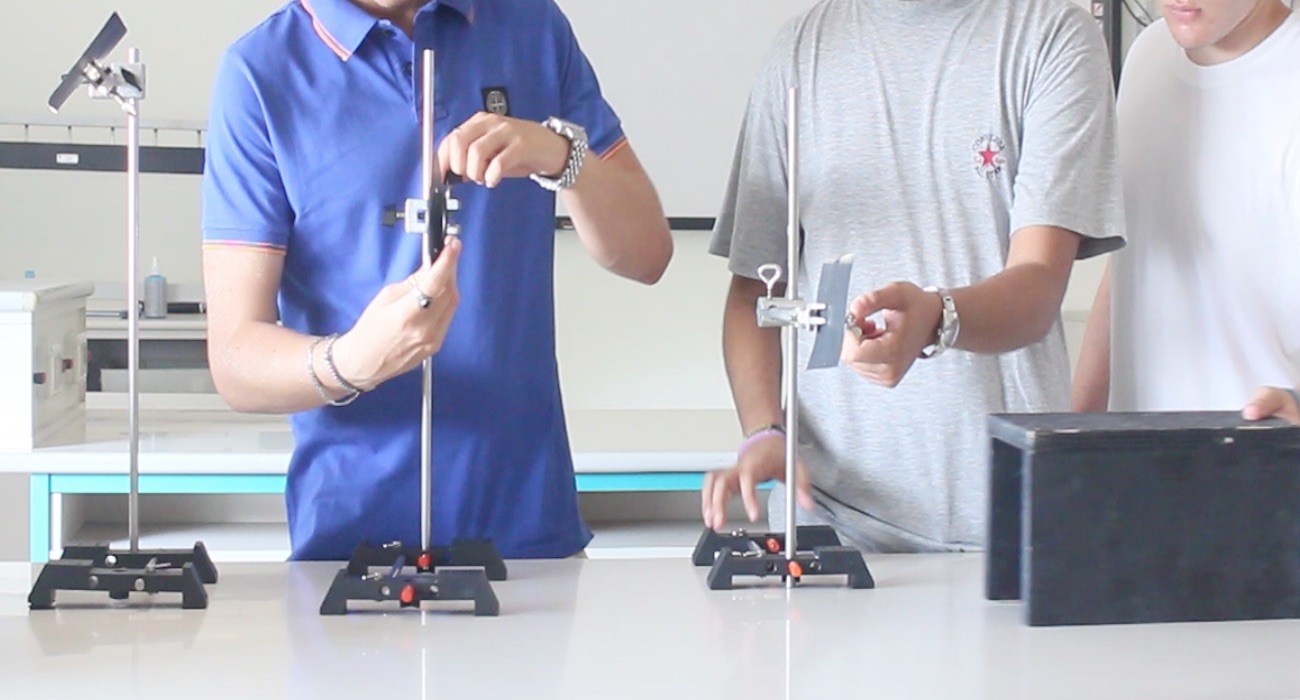} & 
\includegraphics[width=4.7cm]{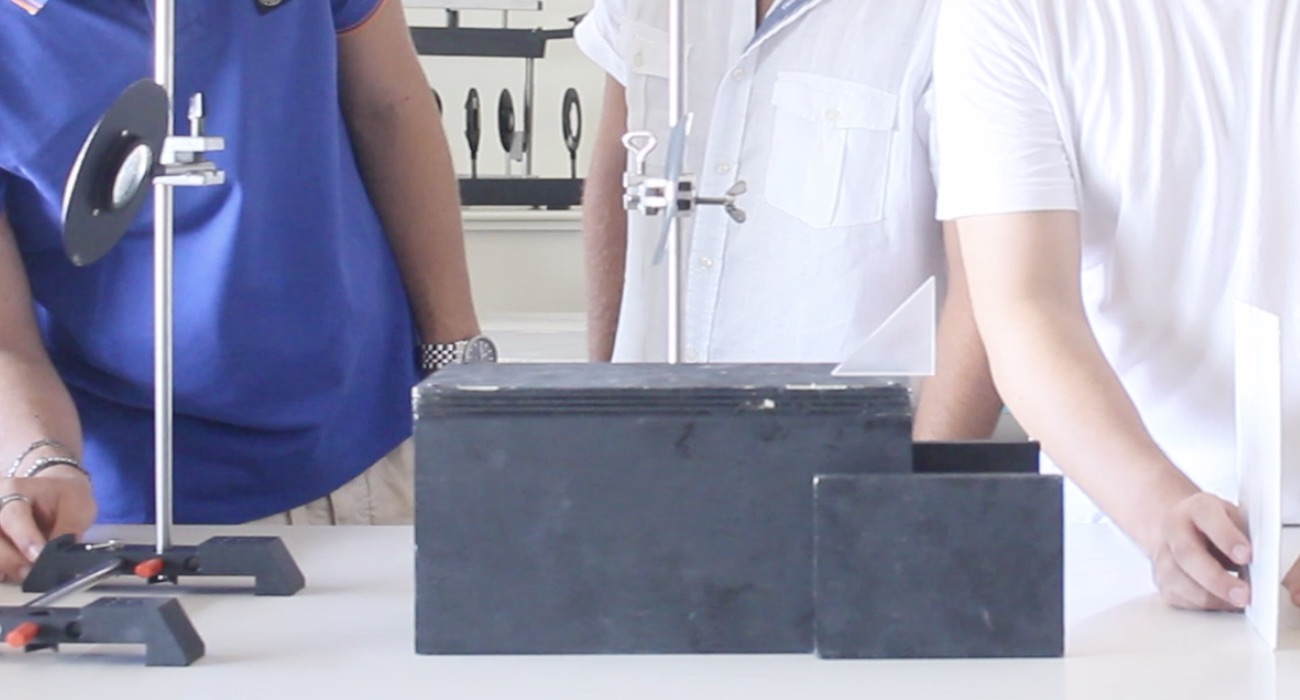} \\
& \includegraphics[width=4.7cm]{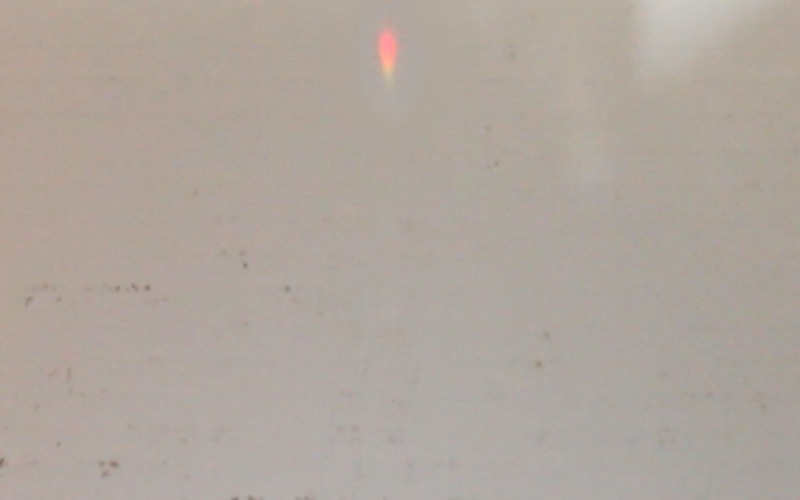} & \includegraphics[width=4.7cm]{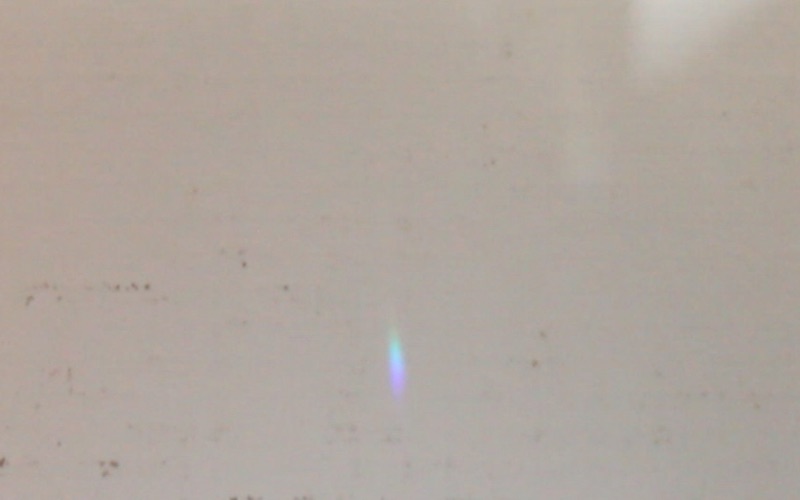} 
\end{tabular} 
\caption{Experiment N.10: one light ray coming from a prism and crossing two small distant holes is refracted by a second prism; by rotating the first prism, the colour selected by the first hole changes (from red to blue), but this light ray refracted by the second one does not change its colour although it is deflected, changing the position of its image on the distant screen (enlarged view in the last two images).}
\label{fig10}}
\end{figure}

\noindent \underline{\it Experiment N.10: experimentum crucis}
\begin{quote}
[Refracted the light through a prism and conveyed through a small hole on a board, it was refracted by a second prism after passing through a hole on a second board.] This done, I took the first prism in my hand, and turned it to and fro slowly about its axis [...]. And I saw by the variation of those places, that the light, tending to that end of the image, towards which the refraction of the first prism was made, did in the second prism suffer a refraction considerably greater than the light tending to the other end. [...] Light consists of rays differently refrangible, which, without any respect to a difference in their incidence, were, according to their degrees of refrangibility, transmitted towards divers parts of the wall (Ref. \cite{Oldenburg}).
\end{quote}
The rather difficult {\it experimentum crucis} was realized with a large glass prism (able to be rotated around its axis) and a smaller plastic prism. The ``single" light ray of given colour (practically, only red and blue) was selected by using two small holes on a paper card placed at a distance each other; an additional converging lens was also used in order to focus light. The coloured light refracted by the second prism was then collected upon a distant white screen (see Fig. \ref{fig10}). The shift in position (but not in the nature of colour) of the refracted image resulted to be quite evident at some large distance.

\section{Conclusions}

\noindent We have here presented an advanced scientific project specially devised for outstanding students of High Schools (or even undergraduate students), with the explicit aim of letting their specific abilities to express to the maximum. Most of (if not all) the projects devoted to enhance students scientific abilities, science popularization and public engagement, indeed, are often prepared for uneducated and (initially) not interested people, so that their basic aim is to capture their attention and stimulate scientific reasoning -- a certainly laudable goal. Little (or no) room is then left to enhance and drive already developed capabilities of students interested in science, with the evident risk to lose the most clever people, or even just to flatten their preparation. In order to fill this gap, we have thus developed and experimented an advanced project along different action lines, centered around a well-known historical case, namely that of Newton's theory of colours. 

The first part of the project was devoted to let the students to {\it think} as Newton (or as any modern science scholar of the western seventeenth century) did, by building step by step all his knowledge and reasoning about the topic (see Sect. 3). The second part was, instead, intended to let the students to {\it work} as Newton did, by performing the whole series of the original experiments described in his manuscripts, and by making recourse to the school lab resources (see Sect. 4). The third part aimed at letting the students to {\it deduce} as Newton did about light and colours directly followed the second one, both in general and specific interpretations and conclusions. Finally, the whole path of Newton's experimentation about prisms, leading to his theory of colours, was asked the students to {\it present} to the general public, especially during two Science Festivals, in order to test also their abilities in communicating what they just learned. 

During any stage of the project, the (few) students were always guided by a teacher who, however, limited his task to coordinate and provoke the students, just by setting the point, and only rarely by addressing the given problem (which, instead, clearly emerged from the original texts, chosen by the teacher). The whole set of Newton's ten experiments, including the famous {\it experimentum crucis}, was also filmed at the end of the project, as an important part of the same (during which a number of unexpected technical topics emerged and were clarified), was presented at the mentioned Science Festivals and is currently visible on the YouTube platform \cite{youtube}. 

The choice of the subject of the present project was certainly motivated by the aim to let the students to build all their knowledge about it from the beginning, without taking anything for granted, during a limited number of months. However, no less a role was played by the purpose to plot a historically informed activity,  given the firm belief that the History of Physics gives an exceptional chance in promoting science even at a deep level.  Last but not least, indeed, the subject of Newton's theory of light and colours still entertains a current research debate (see Ref.s \cite{Takuwa, Grusche} and references therein), although we here present it for the first time (to the best of our knowledge) in connection to didactics and science popularization.

The extremely favourable reception of the project (in all its parts) by the students involved, along with the enormous success granted to them during public events, certainly encourages to propose more advanced projects aimed at forming a well prepared class of future scientists, though without forgetting uneducated people.

% ``

\section*{Acknowledgments}

\noindent The present work would never have seen the light without the fundamental contribution of Daniele Aulitto, Francesco Gasperini, Francesco Granata, Matteo Olimpo and Francesco Panico. Special thanks are also due to Biagio Antonio Coppola for his kind assistance in video making, as well as to SISFA and Naples' Unit of I.N.F.N. for encouraging the activity presented here.


\section*{References}

\begin{thebibliography}{99}

\bibitem{youtube} 
The colours of Newton's {\it Opticks} videos \\
https://www.youtube.com/playlist?list=PLTGvK6jMx5QCxKWsufvf0X6Z6yqAKIQ98

\bibitem{FST} 
Fondazione Scienza e Tecnica http://www.fstfirenze.it

\bibitem{SISFA} 
SISFA -- Italian Society for the History of Physics and Astronomy http://www.sisfa.org/

\bibitem{Weisskopf}
Weisskopf V F  1986 Search for Simplicity: Mountains, waterwaves, and leaky ceilings {\it Am. J. Phys.} {\bf 54} 110-111 

\bibitem{Baeyer}
von Baeyer H C  1993 {\it The Fermi Solution} (New York: Random House)

\bibitem{Weinstein}
Weinstein L and Adam J A 2008 {\it Guesstimation: Solving the World's Problems on the Back of a Cocktail Napkin} (Princeton: Princeton University Press)

\bibitem{Westfall}
Westfall R S 1983 {\it Never at rest: a biography of Isaac Newton} (Cambridge: Cambridge University Press)

\bibitem{Mangio}
Mangio C  1961 Cenni sulle teorie cromatiche dei greci  e loro applicazione architettonica {\it Studi Classici e Orientali} {\bf 10} 214-223

\bibitem{Plato}
Scott D  2005 {\it Plato's Meno} (Cambridge: Cambridge University Press)

\bibitem{Euclid}
Burton H E  1945 The Optics of Euclid {\it J. Opt. Soc. Am.} {\bf 35} 357-372.

\bibitem{Lucretius}
Smith M F 2011 {\it Lucretius: On the Nature of Things} (Indianapolis: Hackett)

\bibitem{Ptolemy}
Mark Smith A 1996 {\it Ptolemy's theory of visual perception} (Philadelphia: The American Philosophical Society)

\bibitem{Avicenna}
McGinnis J  2010 {\it Avicenna} (New York: Oxford University Press)

\bibitem{Alhazen}
El-Bizri N 2006 Ibn al-Haytham or Alhazen, in Meri J W (ed.) {\it Medieval Islamic Civilization: An Encyclopaedia. II} (New York: Routledge)

\bibitem{Grosse}
Sparavigna A C 2013 On the Rainbow, a Robert Grosseteste's Treatise on Optics {\it Int. J. Sciences} {\bf 2} 108-113

\bibitem{Bacon}
Parkhurst C 1990 Roger Bacon on Color, in Selig K L and Sears E (eds.) {\it The verbal and the visual: essays in honor of William Sebastian Heckscher} (New York: Italica Press)

\bibitem{Giudice}
Giudice F  2009 {\it Lo spettro di Newton} (Rome: Donzelli)

\bibitem{Kepler}
Donahue W H 2000 {\it Johannes Kepler: Optics} (Sante Fe: Green Lion Press)

\bibitem{Descartes}
Mark Smith A  1987 {\it Descartes's Theory of Light and Refraction: A Discourse on Method} (Philadelphia: The American Philosophical Society)

\bibitem{GiudiceNewton}
Giudice F  2006 {\it Isaac Newton: Scritti sulla luce e i colori} (Milan: BUR)

\bibitem{NewtonLectiones}
Newton I  1729 {\it Lectiones opticae} (London: Innys)

\bibitem{Oldenburg}
Newton I 1672 New theory about light and colours {\it Philos. Trans. R. Soc. London} {\bf 80} 3075-3087

\bibitem{NewtonOpticks}
Newton I  1704 {\it Opticks} (London: Smith and Walford)

\bibitem{UCDL}
University of Cambridge Digital Library - Newton Papers \\
https://cudl.lib.cam.ac.uk/collections/newton/5

\bibitem{Takuwa}
Takuwa Y 2013 The historical transformation of Newton's experimentum crucis {\it Hist. Sci.} {\bf 23} 113-140

\bibitem{Grusche}
Grusche S 2015 Revealing the nature of the final image in Newton's experimentum crucis {\it Am. J. Phys.} {\bf 83} 583-589

\end{thebibliography}
\end{document}